\documentclass[amsmath, twocolumn, trackchanges]{aastex702}

\defcitealias{2015AA...584A..87C}{CT15}
\defcitealias{2006ApJS..164...81M}{MK06}

\begin{document}

\title{Thermal Radio Fractions in Local Star Forming Galaxies from Total H$\alpha$ Luminosities: Origin of the Non-Linear Radio-FIR Correlation}

\author[orcid=0000-0002-6369-6266]{Tirth D. Surti}
\affiliation{Cahill Center for Astronomy and Astrophysics, California Institute of Technology, 1216 E California Blvd, Pasadena, CA 91125,
USA}
\email[show]{tsurti@caltech.edu}

\author[orcid=0000-0002-7252-5485]{Vikram Ravi} 
\affiliation{Cahill Center for Astronomy and Astrophysics, California Institute of Technology, 1216 E California Blvd, Pasadena, CA 91125,
USA}
\email{}

\author[orcid=0000-0002-6479-6242]{Allison Matthews}
\affiliation{Carnegie Observatories, 813 Santa Barbara Street, Pasadena, CA 91101, USA}
\email{}

\begin{abstract}

The observed tight radio-FIR correlation among star forming galaxies (SFGs) is used to convert 1.4 GHz radio luminosities to star formation rates. The recently observed sub-linear radio-FIR correlation, particularly for radio luminosities below $\log(L_{1.4}/(\mathrm{W\, Hz^{-1}})) = 22.5$, suggests that synchrotron emission must be increasingly suppressed in less radio luminous SFGs, resulting in larger thermal radio fractions relative to ${\sim}L_{\star}$ SFGs. The thermal fraction dependence on total radio luminosity is not well constrained due to the observational difficulty of isolating thermal radio emission in direct radio SED constructions. Due to physically bound origins of H$\alpha$ and thermal radio emission, total dust-corrected H$\alpha$ luminosities provide a promising avenue to estimate thermal radio fractions. Using a sample of 278 SFGs with $19.5\lesssim \log(L_{1.4}/(\mathrm{W\, Hz^{-1}}))\lesssim 22.75$ from the CALIFA DR3 wide-field integral field unit, PHANGS narrow-band H$\alpha$, and Bok 2.3m telescope drift-scan spectra surveys, we estimate a dependence of $f_t^{1.4\,\mathrm{GHz}}\propto L_{1.4}^{-0.34\pm 0.02}$, indicating that the thermal fraction can change by ${\sim}1\,\mathrm{dex}$ over this luminosity range. In order to explain the non-linear radio-FIR correlation, we require FIR emission to underestimate star formation rates relative to thermal radio emission by up to ${\sim}1.9\times$ for SFGs at ${\sim}L_{\star}/10$ under the standard calibration, consistent with previous works that infer lower optical depths to UV emission in less luminous SFGs. We present a cosmic ray loss model that can reproduce the observed thermal fraction dependence on radio luminosity, within which synchrotron suppression in less luminous SFGs is driven by weaker magnetic field strengths leading to more efficient escape.
\end{abstract}

\keywords{\uat{Radio continuum emission}{1340} --- \uat{Star formation}{1569} --- \uat{Cosmic rays}{329} --- \uat{Interstellar medium}{847}}


\section{Introduction}
The radio-infrared (IR) correlation is a tight relationship observed in star forming galaxies \citep{1985ApJ...298L...7H, 2003ApJ...586..794B}, relating the total 1.4 GHz radio emission to the total infrared ($8$--$1000\,\mathrm{\mu m}$, TIR) or far infrared (FIR) emission. FIR luminosities are typically defined as a linear combination of IRAS 60$\mu \mathrm{m}$ and $100\,\mu \mathrm{m}$ luminosities. This correlation is what validates using radio emission as a dust-unbiased star formation tracer and has consequently been used to convert radio luminosities into star formation rates in deep radio surveys \citep[e.g.,][]{Matthews_2021, 2022ApJ...941...10V, 10.1093/mnras/stag616}. The total radio emission in star forming galaxies (SFGs) is generally comprised of thermal emission from active star forming HII regions and the diffuse warm-ionized medium (WIM) and non-thermal emission from cosmic rays accelerated by supernovae \citep{1992ARA&A..30..575C}. Direct radio SED constructions of nearby Milky Way-like spiral galaxies have shown that at 1.4\,GHz, the fraction of thermal radio emission is ${\sim}10\%$ \citep{1990ApJ...357...97C, 2003ApJ...586..794B, 2017ApJ...836..185T}. The TIR and FIR emission trace the reprocessed optical and Balmer/Lyman continuum emission from active star formation by dust. Defining the radio-FIR correlation as
\begin{equation}
q_{\mathrm{FIR}}=\log\left(\frac{L_{\mathrm{FIR}}/(3.75\times10^{12}\,\mathrm{Hz})}{L_{1.4}}\right),
\end{equation}
\cite{Murphy_2009} show that by relating the thermal radio emission to TIR emission under the assumption that they trace the same star formation rate as
\begin{equation}
L_{\nu}^{\mathrm{T}}\approx 6.6\times10^{-17}\left(\frac{T_e}{10^{4}\,\mathrm{K}}\right)^{0.45}\left(\frac{\nu }{\mathrm{GHz}}\right)^{-0.1}L_{\mathrm{TIR}},
\end{equation}
the radio-FIR correlation can be rewritten as a function of the thermal radio fraction at 1.4\,GHz ($f_t$),
\begin{equation}
    q_{\mathrm{FIR}} = q_0+\log(f_t),
\end{equation}
under a constant $L_{\mathrm{TIR}}/L_{\mathrm{FIR}}$ ratio. For an electron temperature $T_e\approx 10^{4}\,\mathrm{K}$, $q_0\approx 3.32$ is consistent with the observed value of $\langle q_{\mathrm{FIR}}\rangle \approx 2.34\pm 0.26\,\mathrm{dex}$ \citep{2001ApJ...554..803Y} given an average $10\%$ thermal fraction in star forming galaxies and a local $\langle L_{\mathrm{TIR}}/L_{\mathrm{FIR}}\rangle\approx 2$ \citep{2003ApJ...586..794B, 10.1093/mnras/staf1536}.

\cite{2003ApJ...586..794B, 2010ApJ...717....1L} argue, however, that less luminous/massive SFGs (${\lesssim}L_{\star}$) are optically thin to UV emission, resulting in a smaller fraction of UV emission being reprocessed into FIR. This results in FIR emission underestimating star formation rates in such SFGs. In order to keep the radio-FIR correlation approximately linear, there must be a conspiracy that the amount of non-thermal emission in SFGs fainter than the Milky Way must also be suppressed by up to ${\sim}2$--$3\times$, which would otherwise cause a decrease in the observed $q_{\mathrm{FIR}}$. Furthermore, recent work by \cite{Matthews_2021, 2021A&A...647A.123D, 10.1093/mnras/stab746, 10.1093/mnras/staf1536} has shown that the radio-FIR correlation is sublinear. That is, $q_{\mathrm{FIR}}$ likely rises with lower radio luminosities for $\log(L_{1.4}/(\mathrm{W\, Hz^{-1}}))\lesssim 22.5$, indicating that the amount of synchrotron suppression may be larger than implied by a linear radio-FIR correlation. However, given the limited number of radio luminous SFGs and different galaxy samples, it is currently unclear whether $q_{\mathrm{FIR}}$ continues to decrease \citep[e.g.,][]{10.1093/mnras/stab746, 10.1093/mnras/staf1536} or flatten \citep[e.g.,][]{Matthews_2021} for $\log(L_{1.4}/(\mathrm{W\, Hz^{-1}}))\gtrsim 22.5$.

The amount of synchrotron suppression on average and its trend with radio luminosity, however, is not well known due to the lack of thermal fraction measurements in a large sample of SFGs, especially for $L_{1.4}\ll L_{\star}\approx 2\times10^{21}\,\mathrm{W\, Hz^{-1}}$ \citep{2026arXiv260602717S}. The relative contributions of synchrotron suppression and IR emission underestimating star formation rates in driving the non-linear radio-FIR correlation at low radio luminosities can be used to understand dust content differences in galaxies and place constraints on ISM and cosmic ray loss physics as a function of radio luminosity.

Synchrotron suppression in less luminous and dwarf SFGs has been interpreted to arise from their inability to confine cosmic rays given their generally weaker magnetic field strengths \citep[e.g.,][]{2010ApJ...717....1L, 2011A&A...529A..94C, 2012MNRAS.423L.127R, 2018ApJS..234...29H} and are thus not good cosmic ray calorimeters. An alternative interpretation is that some SFGs have sufficiently young starbursts that $M_{\star}>8\,\mathrm{M_{\odot}}$ stars have not yet had enough time to undergo core-collapse supernovae to produce a normal cosmic ray population \citep[e.g., nascent starbursts][]{2003ApJ...593..733R, 2006ApJ...646..841R, 2006A&A...460...67H} but are actively producing thermal radio emission via HII regions.

While there is some evidence that lower radio or IR luminosity SFGs in the local universe have higher thermal fractions \citep[e.g.,][]{2003ApJ...586..794B, 2012MNRAS.423L.127R, 2017ApJ...836..185T, 2018ApJS..234...29H, 2018A&A...611A..55K}, a systematic study over a large sample of SFGs through direct SED construction can be expensive. Constraining robust thermal fractions through radio SED fitting \citep[e.g., as done by][]{2017ApJ...836..185T, 2018A&A...611A..55K, 10.1093/mnras/stx2613} requires obtaining flux density measurements from $\lesssim 1\,\mathrm{GHz}$, where synchrotron dominates, out to $\gtrsim 33\,\mathrm{GHz}$, where thermal emission starts to dominate in typical SFGs. This encompasses over an order of magnitude change in resolution with a given radio telescope, which necessitates using different radio telescopes to fill out the SED, for example, to avoid confusion at low frequencies or resolving out diffuse, low surface-brightness thermal emission at high frequencies. Additionally, constraints on thermal radio fractions from radio SEDs can be model dependent, requiring choices on the shape of the non-thermal synchrotron component as either a strict power law or having a cutoff due to cosmic ray losses \citep[as explored in detail in][]{2018A&A...611A..55K}.

Instead of direct radio SED constructions, total dust-corrected H$\alpha$ luminosities (either via the Balmer decrement or via mid-IR emission) provide a promising avenue to estimate thermal fractions for a large sample of SFGs, as has frequently been done before \citep[e.g.,][]{2012MNRAS.423L.127R,2017ApJ...836..185T, 2018ApJS..234...29H, 10.1093/mnras/sty3199, Tabatabaei_2025}. Both H$\alpha$ and thermal radio luminosities are directly related to the emission measure through the WIM and hence the cumulative ionizing photon rate from star forming regions assuming photoionization equilibrium. Their ratio only depends on the electron temperature and the frequency of observation as
\begin{equation}
    \frac{S_{\nu}^{T}}{S_{H\alpha}} = \frac{1.25}{10^{14}\,\mathrm{Hz}}\left(\frac{T_e}{10^{4}\,\mathrm{K}}\right)^{0.59}\left(\frac{\nu}{\mathrm{GHz}}\right)^{-0.1},
\end{equation}
adapted from \cite{1986A&A...155..297C, 1992ARA&A..30..575C}, assuming a typical electron density $n_e=100\,\mathrm{cm^{-3}}$ and Case B recombination, valid for temperatures between $5000$ and $20000\,\mathrm{K}$ and under photoionization equilibrium. While even dust-corrected H$\alpha$ may underestimate the total ionizing photon rate due to Lyman continuum absorption by dust in higher metallicity regions \citep{Inoue_2001, 2003ApJ...583..727D} or escape in lower metallicity regions \citep{10.1111/j.1365-2966.2012.21107.x}, the thermal radio emission is affected by the same losses, so the ratio normalizes out differences in ISM conditions between different SFGs. 

Depending on the method to identify it, a large fraction of H$\alpha$ emission (${\sim}40$-$60\%$) is expected to come from the diffuse ionized gas \citep[DIG;][]{2007ApJ...661..801O, belfiore2022}, which has been inferred to be produced from leaky Lyman continuum photons ionizing neutral gas between HII regions \citep{1996AJ....111.2265F, 2009RvMP...81..969H}. We note that other possible candidates can contribute to H$\alpha$ emission in the DIG such as hot low-mass evolved AGB stars \citep[HOLMES,][]{10.1111/j.1365-2966.2011.18848.x, 2011RMxAC..40...96S} or low-velocity shocks \citep{1981MNRAS.197..659P, Monreal-Ibero_2006}. Recent observations from the PHANGS-MUSE survey \citep{belfiore2022}, however, suggest a small (${\sim}2\%$) contribution to the H$\alpha$ luminosity in the DIG by HOLMES. Hence, modulo any lost ionizing photons, \cite{belfiore2022} indicate that integrated H$\alpha$ or H$\alpha$ measured on angular scales larger than HII regions is still a reliable star formation rate tracer based on standard calibrations.

As with direct SED constructions, using dust-corrected H$\alpha$ is not without complications. \cite{Dong_2011} suggest that some H$\alpha$ emission in the DIG may come from gas not in photoionization equilibrium, in which case Equation 4 does not hold or must be normalized to a lower effective electron temperature. Nevertheless, H$\alpha$ luminosities have been shown to produce thermal fractions consistent with those inferred from direct SED constructions in local SFGs. For example, among the sample of normal SFGs analyzed in \cite{2017ApJ...836..185T}, median thermal fractions estimated at $5\,\mathrm{GHz}$ from dust-corrected H$\alpha$ luminosities based on 24$\,\mathrm{\mu m}$ emission originally calibrated to Balmer-decrement-corrected H$\alpha$ luminosities \citep{2009ApJ...703.1672K} tend to agree with those derived from radio-SED fitting. The typical absolute difference between the two approaches is just ${\sim}5.0\%$ with standard deviation of $4.4\%$. Additionally, \cite{2011ApJ...737...67M} find close agreement between SFRs estimated using H$\alpha$ and thermal emission at 33 GHz (independently) in extranuclear star-forming regions of NGC 6946. Here, H$\alpha$ was corrected for dust based on $24\,\mathrm{\mu m}$ emission calibrated to extinction-corrected Pa$\alpha$ emission \citep{2007ApJ...666..870C}.

Additionally, it is possible that individual HII regions may have sufficiently high emission measures that at 1.4 GHz, thermal radio emission becomes optically thick with $L_{\nu}^{T}$ tracing the Rayleigh-Jeans black-body emission. This has been observed in dense starburst regions \citep[e.g.,][]{2010MNRAS.405..887C} at GHz frequencies. This causes H$\alpha$ emission to over-estimate thermal radio emission. However, a large fraction of thermal radio and H$\alpha$ emission is expected to come from the DIG \citep{2007ApJ...661..801O, belfiore2022} thus reducing average emission measures through the full galaxy line of sight \citep{2015AJ....149...32M}. Additionally, more recent LOFAR observations suggest that spectral curvature observed at low frequencies $<1\,\mathrm{GHz}$ in typical SFGs is independent of inclination \citep[which would otherwise affect the emission measure,][]{2018A&A...619A..36C}, suggesting that spectral curvature is mainly due to cosmic ray losses rather than free-free absorption. Thus, for integrated H$\alpha$ measurements among our sample of SFGs, we do not expect free-free absorption to be significant at 1.4 GHz.

%The steeper non-linear relationship found between total 1.4 GHz radio and dust-corrected H$\alpha$ luminosities by \cite{2009ApJ...703.1672K} than between TIR and dust-corrected H$\alpha$ luminosities was interpreted to be due to radio emission declining more steeply than dust emission in less luminous SFGs, hinting at changing cosmic ray lifetimes as a function of luminosity in these SFGs. However, an explicit connection to the thermal fraction dependence on radio luminosity was not made. This steeper relationship of dust-corrected H$\alpha$ luminosities against 1.4 GHz radio luminosities can be used to explain the non-linear radio-FIR correlation that has recently been established.

In this paper, we aim to clarify (i) the extent of synchrotron suppression as a function of the total 1.4 GHz radio luminosity for normal SFGs statistically, (ii) directly determine the relative role of synchrotron suppression versus IR emission underestimating star formation rates in order to explain the non-linear radio-FIR correlation, and (iii) investigate the cosmic ray loss mechanisms responsible for producing a systematic change in thermal radio emission as a function of total 1.4 GHz radio luminosity. To do so, we use the wide field-of-view (FOV) integral field unit CALIFA \citep{2012A&A...538A...8S, 2014A&A...569A...1W}, Bok 2.3m telescope drift-scan spectra \citep{2006ApJS..164...81M}, and  PHANGS-H$\alpha$ \citep{2026A&A...710A.412R} surveys to directly constrain the integrated H$\alpha$ emission and 1.4 GHz thermal fractions in local SFGs having radio luminosities $19.5\lesssim \log(L_{1.4}/(\mathrm{W\, Hz^{-1}}))\lesssim 22.75$, a luminosity range over which the radio-FIR correlation is securely non-linear. In Section 2, we describe the SFGs selected from the CALIFA, Bok telescope drift-scan, and PHANGS surveys and how AGN were filtered out. In Section 3 we describe the total H$\alpha$ luminosity fitting procedure for the CALIFA galaxies and the corresponding aperture corrections made in cases where the IFU does not cover the entire spatial extent of the SFG. In Section 4, we describe aperture corrections for the drift-scan spectra. In Section 5, we present the results of the thermal fraction dependence on the total radio luminosity (addressing (i)). In Section 6, we discuss the systematic uncertainties that could affect the derived thermal fractions. Finally, in Section 7, we discuss the origin of the non-linear radio-FIR correlation and its scatter in terms of the observed thermal fraction relation (addressing (ii)), and the origin of the synchrotron suppression using a basic steady-state cosmic ray loss and injection model (addressing (iii)).

We assume a convention $S_{\nu}\propto \nu^{\alpha}$ with $\alpha<0$ to represent the generally decreasing radio spectrum of SFGs with frequency at GHz frequencies.

\section{Data Selection}

\subsection{CALIFA DR3}
Our sample of SFGs for H$\alpha$ line fitting is selected from the Calar Alto Legacy Integral Field Area Survey \citep[CALIFA][]{2012A&A...538A...8S} Data Release 3 \citep[DR3,][]{2016A&A...594A..36S}, within which the ``mother sample'' (MS) of ${\sim}600$ galaxies were selected from the Sloan Digital Sky Survey \citep[SDSS DR7,][]{2009ApJS..182..543A} having redshifts $0.005 < z<0.03$ and sizes $45^{\prime\prime} < \mathrm{isoA_r} < 79.2^{\prime\prime}$ \citep{2014A&A...569A...1W}. $\mathrm{isoA_r}$ is the isophote major axis of the galaxy in SDSS $r$-band representing $\mu_r=25\,\mathrm{mag/arcsec^2}$. Since the original survey description, additional galaxies have also been added into an ``extension sample'' (ES). The survey was designed so that most galaxies could be covered out to at least the SDSS optical isophotal radius by the Calar Alto observatory PPak integral field unit (IFU), which has a field of view of ${\sim}1\,\mathrm{arcmin^2}$. Therefore, the CALIFA IFUs are generally expected to capture a large fraction of the extended H$\alpha$ emission. DR3 contains 646 data cubes (denoted ``v500'') of galaxies in both the MS and ES at a median resolution of $R{\sim}850$ over a wavelength range of $3750$--$7500$\, \AA, which is sufficient to cover the H$\beta$--[O\,III] and [N\,II]--H$\alpha$ line complexes for recovering dust-corrected H$\alpha$ luminosities. The typical angular resolution of the observations is at $2.5^{\prime\prime}$ FWHM. The reduced data cubes are already corrected for foreground Galactic extinction using the \cite{1998ApJ...500..525S} dust maps and the \cite{1989ApJ...345..245C} extinction curves for $R_V=3.1$.
%Following DR3, an additional ${\sim}200$ IFUs have been since been included in a remastered release \citep{10.1093/mnras/stad3119} into the extended sample under a different data reduction procedure, but we were made aware of this dataset after much of the analysis was already done on the DR3 datacubes. Based on the following selection criteria, we expect ${\sim}30$ more SFGs from this remastered sample that were left out of this analysis with similar radio luminosities and are therefore unlikely to change the conclusions of this work.

Galaxies in this sample are morphologically heterogeneous, including active mergers, quiescent elliptical galaxies, SFGs, and AGN. Therefore, the original sample must be carefully filtered to select only those that have minimal evidence of AGN or merger-induced activity. Furthermore, given the large angular extents of these SFGs, reliable radio flux densities at 1.4 GHz require cross matches with sensitive, low resolution surveys, such as the NRAO VLA Sky Survey \citep[NVSS,][]{1998AJ....115.1693C}.

To narrow the sample of galaxies from CALIFA DR3, we first select those that appear isolated. Galaxies in active mergers may have excess recombination line emission not from star formation due to cooling shocked gas \citep[e.g.,][]{Rich_2015}. Tables provided in the CALIFA DR3 data release\footnote{\url{https://califa.caha.es/dr3-tables.html}} provide by-eye morphological classifications, which are used to filter out the 49 galaxies that appear to be in active mergers. We then only select galaxies that appear to have NVSS cross-matches based on the nearby radio-emitting galaxy sample of \cite{2019ApJ...872..148C}. While this sample is restricted to IR-bright sources in the local universe (having $k_{\mathrm{20fe}}\leq 12.25$), cross matching to this sample over the original NVSS catalog has two significant benefits: (i) WISE colors and $q_{\mathrm{FIR}}$ have already been used to classify the radio sources as likely SFG or AGN, and (ii) careful analysis has been done to identify confusion and cross-match sources with 2MASX counterparts.

In addition to the IR-selection performed by \cite{2019ApJ...872..148C}, we select objects consistent with star formation or starbursts based on optical line ratios in the BPT diagram \citep{1981PASP...93....5B}. Nuclear classifications were already made by \cite{Jonic2021} for the 483/542 galaxies in the MS that have detectable emission lines. Out of the 483 with nuclear optical classifications, we select only those with NVSS cross matches in \cite{2019ApJ...872..148C}, within which they are also classified as an SFG based on IR emission, and that are classified either as ``star-forming'' or ``intermediate'' based on the optical line ratios. The intermediate classification encompasses galaxies that are still consistent with the maximum starburst line of \cite{2001ApJ...556..121K}. This results in 152 galaxies consistent with star formation or intermediate from the MS.

We identified 23 additional galaxies in the ES and 1 galaxy in the MS (NGC 2253) that could be cross matched to NVSS and classified by \cite{2019ApJ...872..148C} as star forming but whose nuclear optical classifications are not known. We proceeded to extract nuclear spectra for these 24 galaxies to identify the corresponding ionizing source.

\subsubsection{Remaining Sample Classification}
Of the unclassified 24 galaxies, we extract the spectrum in a $3^{\prime\prime}\times3^{\prime\prime}$ region in the photometric center, consistent with the FWHM angular resolution of the CALIFA IFUs. The photometric center is identified to be the brightest spatial pixel in the wavelength-averaged image after masking foreground sources and bad spectral pixels as identified by the provided ``BADPIX'' cube. We then obtain the averaged spectrum in the extracted region, propagating the error in each spectral pixel accordingly. We note that NGC 3655 (having a CALIFA name of SN2002ji) has a bright foreground cataclysmic variable star adjacent to the core (AT 2018aqc) which is masked, so only a portion of the core spectrum is represented in the extraction. No selected source shows clear evidence of broad-line AGN emission in the observable Balmer emission lines.

For each extracted spectrum, we use the package \textsc{PyPipe3D} \citep{2022NewA...9701895L} to first fit the stellar continuum. We utilize the provided MILES stellar template library for the continuum fits \citep{2006MNRAS.371..703S, 2011A&A...532A..95F}, which we find to give an acceptable continuum subtraction around H$\beta$, which often exhibits nebular emission superimposed onto continuum absorption. We fit the continuum over the fully unvignetted wavelength range of $4240$--$7140$\,\AA. We adopt the median spectral resolution of $6$\,\AA\, FWHM, so in most cases the emission line velocities are not well-resolved. In the continuum fitting, we mask common sky-lines that lead to poor sky-subtraction with half-width 12\,\AA\, (twice the spectral resolution), the most prominent being [O\,I]$\lambda$5577 \AA. SFG emission lines are masked internally. We use initial systemic velocity guesses supplied by the tables in the CALIFA DR3 data release. For the H$\beta$--[O\,III] complex, we fit over rest wavelengths $4800$-$5050$\,\AA, and for the [N\,II]--H$\alpha$ line complex, we fit over the rest frame range of $6510$--$6640$\,\AA, sufficient to cover the width of the lines and establish the continuum-subtracted baseline level. In the emission line fits, we include an additional constant normalization (through PyPipe3D's \textsc{poly1d}) to account for any remaining flux from the continuum level. In general, we find that the normalization fitted is a negligible fraction of the observed line peak intensities, suggesting acceptable continuum subtraction.

\begin{figure}
    \centering
    \includegraphics[width=\linewidth]{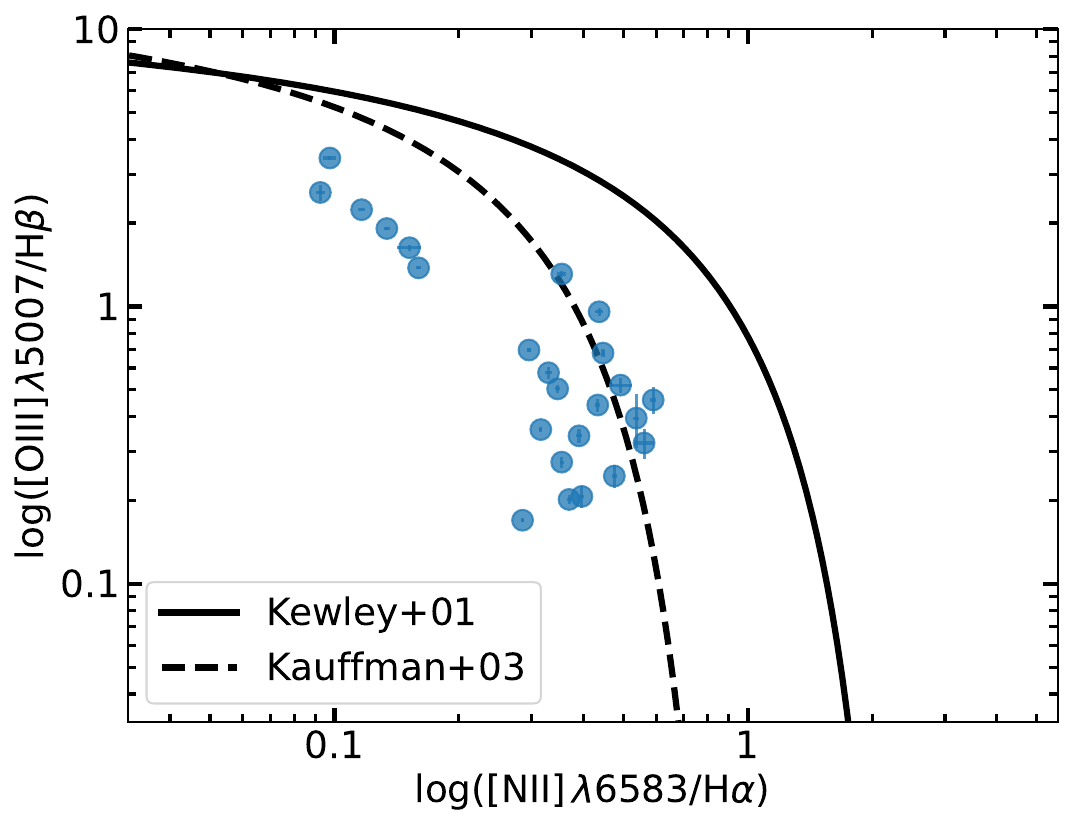}
    \caption{Placement of the 24 optically unclassified galaxies on the BPT diagram based on the average central $3^{\prime\prime}$ spectrum. All galaxies here are consistent with star forming or ``intermediate.''}
    \label{fig:fig1}
\end{figure}

In Figure \ref{fig:fig1}, we show the BPT diagram classifications of the 24 objects, adding the star formation and maximum starburst demarcation lines from \cite{2003MNRAS.346.1055K} and \cite{2001ApJ...556..121K}, respectively. All of the objects are consistent with either star forming or intermediate, being well below the maximum starburst line. This results in a total of 176 NVSS crossed-matched CALIFA galaxies classified both optically and in the NIR as consistent primarily with star forming that we consider for total H$\alpha$ luminosities.

Due to the low observed redshifts of the CALIFA SFGs, they are likely subject to peculiar velocities. Querying the NASA Extragalactic Database (NED), we do not find any redshift-independent distances for the CALIFA SFGs arising from the tip of the red giant branch (TRGB), cepheid variable star, or brightest star methods. We therefore use the heliocentric velocities $cz$ cataloged by \cite{2019ApJ...872..148C} converted to local sheet frame velocities \citep{2008ApJ...676..184T} and the Cosmicflows-4 \citep[CF4,][]{Tully_2023} distance velocity calculator to estimate distances. The distance calculator uses a combination of the elliptical galaxy fundamental plane relation, spiral galaxy Tully-Fisher relation, Type 1a supernovae, cepheids, and TRGB measurements to calibrate peculiar velocities. 

\subsection{Bok Telescope Drift-Scan Integrated Spectra}
The nearby SFG sample of \cite[][hereafter, MK06]{2006ApJS..164...81M} consists of a large number of cross matches with the \cite{2019ApJ...872..148C} SFG sample that are not covered by the recent CALIFA and PHANGs surveys. \citetalias{2006ApJS..164...81M} obtain integrated drift-scan spectra over regions extending out to several arcminutes using the 2.3m Bok Telescope Boller and Chivens Spectrograph. The spectra are obtained at a median spectral resolution of ${\sim}8$\AA\, and apertures are selected to minimize foreground stellar contamination. The fitted spectra are corrected using the \cite{1998ApJ...500..525S} dust maps with the \cite{1994ApJ...422..158O} extinction curve with $R_V=3.1$. We find 218 sources cross-matched with \cite{2019ApJ...872..148C} consistent with the star formation classification, with 11 already picked out by the CALIFA survey, which we drop.

We then query the nuclear optical line flux catalog \citetalias{2006ApJS..164...81M} taken for 142 galaxies in their sample along with SDSS DR20 \citep{2026arXiv260726149S} and DESI DR1 \citep{Abdul_Karim_2026} nuclear spectroscopy, ensuring that the optical center lies within the fiber. We manually inspect any spectra that are flagged as having broad line emission, resulting in only one source (MRK 0863) as having a clear broad component to the H$\alpha$ and H$\beta$ emission. We only select SFGs whose optical line emission is consistent with star formation or intermediate to within $3\sigma$ across all duplicate nuclear measurements, which results in 104 target SFGs.

Following the CALIFA DR3 sample, through visual inspection, we filter out any additional SFGs that show clear evidence of being in an active merger using Pan-STARRS1 cutouts \citep{Flewelling_2020}, of which we find 19. We finally drop any SFGs (10) that were flagged to be observed in non-photometric conditions, which may indicate poor flux calibration. This results in a final clean SFG dataset of 75 SFGs for which we consider total dust-corrected H$\alpha$ luminosities.

Most distances in \citetalias{2006ApJS..164...81M} use the \cite{2000ApJ...529..786M} multi-attractor model and are not redshift-independent. Similar to the CALIFA SFGs, we query NED for cepheid, TRGB, and brightest star distance estimates first. We find 11 NED-based distances from these methods and 1 redshift-independent distance from \citetalias{2006ApJS..164...81M} for a cepheid-based group distance. These distance estimates, being $\lesssim 20\,\mathrm{Mpc}$, can be quite different from the CF4 distance estimates, with a mean absolute percent difference of ${\sim}25\%$. However, the mean \textit{signed} difference is just ${\sim}5\%$, so we do not expect distance uncertainties to on average affect the overall trends derived in this manuscript. Two sources in NED have multiple brightest star distances that are consistent to within ${\sim}20\%$ of each other; we adopt their reported means. The remaining distances are calculated using CF4 using the \cite{2019ApJ...872..148C} heliocentric $cz$ velocities, as with the CALIFA SFGs.

Similarly, we use the measured $cz$ velocities cataloged by \cite{2019ApJ...872..148C} to shift the H$\alpha$ and H$\beta$ line fluxes to be adjusted to the \cite{1989ApJ...345..245C} $R_V=3.1$ extinction curve based on the \cite{1998ApJ...500..525S} dust map $E(B-V)$. This changes the H$\beta$ line flux by at most ${\sim}1\%$ and the H$\alpha$ line flux by even less.

\subsection{PHANGS Narrow-band H$\alpha$}
The PHANGS-H$\alpha$ sample is a subset of 65 SFGs from PHANGS parent sample of $D<20\,\mathrm{Mpc}$ massive SFGs with $\log(M_{\star})>9.75$ that largely fall on the local star formation main sequence. This survey overlaps with the PHANGS-MUSE survey conducted on 19 SFGs in the parent sample \citep{2022A&A...659A.191E}, which, similar to CALIFA obtained large-FOV mosaiced IFU spectroscopy. The narrow-band H$\alpha$ observations were conducted on the MPG 2.2m telescope at the La Silla Observatory and the du Pont 2.5m telescope at the Las Campanas Observatory with a typical seeing ranging from ${\sim}0.6^{\prime\prime}$ to ${\sim}1.5^{\prime\prime}$. The narrowband images are flux-calibrated using color-color relations from nearby Gaia stars and continuum subtracted based on the broadband image after removing the nebular contribution from the broadband flux. Finally, the contamination of [N\,II] within the narrow band image is corrected by using the mass-metallicity and N2-metallicity relations, which can be used to predict the ratio of the [N\,II] to H$\alpha$ flux within the narrow band image. Comparisons to the PHANGS-MUSE dataset is used for verification of the narrow-band H$\alpha$ fluxes.

Similar to the CALIFA galaxies, we were able to cross-match 30 of the 65 PHANGS-H$\alpha$ galaxies to NVSS galaxies in the \cite{2019ApJ...872..148C} sample, noting that many of the PHANGS galaxies are largely in the southern sky and are thus not covered by the NVSS footprint. Two of these are duplicates with the drift-scan, so we drop them in favor of having dust-corrected H$\alpha$ luminosities. For luminosity calculations, to remain consistent with the PHANGs surveys, we use the redshift-independent distances provided by \cite{2026A&A...710A.412R} originally gathered in \cite{2021MNRAS.501.3621A}. Two SFGs have two H$\alpha$ luminosity measurements from the different observations separated by a factor of ${\sim}2$ for which we use the mean in log-space.

To reduce systematics with the CALIFA SFGs, we adjust the Galactic foreground extinction correction to the \cite{1998ApJ...500..525S} and \cite{1989ApJ...345..245C} extinction law using the measured \cite{2019ApJ...872..148C} velocities $cz$. This results in a maximum change in the H$\alpha$ luminosities by a marginal ${\sim}3\%$.

Hence, combined with the CALIFA and \citetalias{2006ApJS..164...81M} galaxies, we consider a total of 278 SFGs, which are summarized in Table \ref{tab:tab1}. 

\section{H$\alpha$ Luminosity Fitting of CALIFA Galaxies}
\cite{2015AA...584A..87C}, hereafter \citetalias{2015AA...584A..87C}, have already performed total H$\alpha$ luminosity fitting on the subset of selected CALIFA galaxies part of the DR2 release, while those released as a part of DR3 still require fitting. Given that the optical (and UV) extent of some galaxies extends beyond the CALIFA IFU FOV, \citetalias{2015AA...584A..87C} performed aperture corrections based on the ratio of the GALEX NUV flux inside and outside of an elliptical aperture with semi-major axis $r=36^{\prime\prime}$ centered on the galaxy. This assumes that extended NUV emission should also closely trace the extended H$\alpha$ and H$\beta$ emission in star forming regions. They use axial ratios derived from B-band surface brightness isophotal photometry to set the minor axis of the elliptical aperture. In this work, we sum the spectrum within a \textit{circular} $r=36^{\prime\prime}$ aperture and thus consider NUV aperture corrections using such aperture. While this may introduce additional noise from off-source spaxels when fitting for H$\alpha$ luminosities, this method is insensitive to asymmetric nebular and UV emission that broad-band optical surface brightness isophotal photometry may not exhibit due to tracing a smoother underlying stellar continuum. It is also insensitive to the possibility that the CALIFA IFU is not centered precisely on the photometric center of the galaxy. As later shown in Sections 4 and 5, our Balmer decrements and thermal fraction estimates are generally consistent with those of \citetalias{2015AA...584A..87C} post aperture-corrections.

Some SFGs in the DR2 sample were not identified to have NUV observations, so their aperture correction factors were estimated based on the empirical relation between the NUV correction factor and the optical size. For robustness, we only adopt total H$\alpha$ measurements from \citetalias{2015AA...584A..87C} that already have NUV magnitudes and therefore direct aperture corrections (total 73), which leaves 103 CALIFA galaxies to identify NUV emission for aperture corrections and manually fit H$\alpha$ luminosities in this work. Table \ref{tab:tab1} identifies which specific CALIFA galaxies were not fitted in this analysis and adopted from previous works. In the remaining sample, we only select those that have NUV cross matches so that the aperture correction factor can be derived.

\subsection{NUV Cross Matching and Aperture Corrections}
For each galaxy, we first cross match to the asymptotic NUV magnitude catalogs of \cite{2015ApJS..220....6B} containing 4138 nearby galaxies with diameters $>1^{\prime}$ and \cite{2018ApJS..234...18B} containing 1931 galaxies at $z<0.01$. The authors obtain asymptotic magnitudes by masking foreground sources and computing the magnitudes within increasingly larger aperture radii $m(<r).$ Both catalogs estimate the asymptotic magnitude by finding the y-intercept to the linear fit on the enclosed aperture magnitude $m(<r)$ versus the gradient $dm/dr$ as done originally by \cite{2001ApJS..133..321C}. If a galaxy is found in both catalogs, we adopt the asymptotic magnitude from \cite{2018ApJS..234...18B}, noting that both catalogs use a consistent UV extinction law with $R_V=3.1$ and $A_{\mathrm{NUV}}=8.0E(B-V)$ from \cite{1989ApJ...345..245C}. We note that for the eight sources considered that are common to both catalogs, the median differences are $0.04\,\mathrm{mag}$ with a maximum of $0.24\,\mathrm{mag}$, thus indicating the consistency between both approaches. 

\begin{figure}
    \centering
    \includegraphics[width=\linewidth]{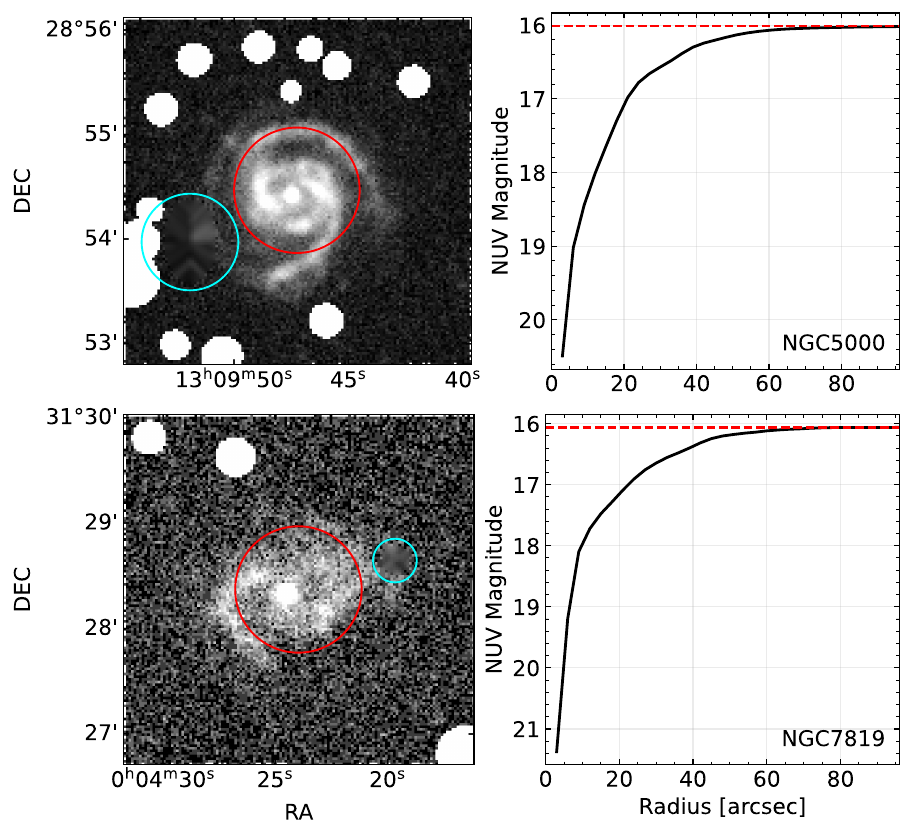}
    \caption{Examples of manual asymptotic magnitude calculations on two sources. The left panels show the CALIFA $36^{\prime\prime}$ aperture (red) centered on the IFU coordinates for comparison to the typical NUV emission extent, and for UGC 04029, we show the smoothed mask-filled interpolated aperture (cyan) due to foreground emission coincident with the galaxy. The right panels show the flattening in the growth in the NUV magnitude as increasingly larger aperture sizes are used, limiting to the asymptotic magnitude (dotted red).}
    \label{fig:fig2}
\end{figure}

Only 43 sources have cross matches to the literature catalogs (within $10^{\prime\prime}$), so for the remaining sources in the GALEX NUV footprint, we manually compute asymptotic magnitudes. Sources that have contaminating foreground emission over the host galaxy or within the $r=36^{\prime\prime}$ aperture are masked using the package \textsc{maskfill} \citep{vanDokkum_2024}, which uses edge pixel information to extrapolate iteratively inwards using a median filter. Other foreground sources are masked. We adopt the standard $3\times 3$ pixel median filter and a final $3\times 3$ pixel boxcar to smoothen the interpolation. Asymptotic magnitudes are calculated using a curve-of-growth analysis on minimum $150^{\prime\prime}$ cutouts with an aperture step size of $3^{\prime\prime}$ centered on the CALIFA cube, performing a linear fit on the last six aperture measurements entirely enclosed in the cutout. We extend the cutout in steps of $50^{\prime\prime}$ if the SFG angular extent is larger than the nominal cutout size. We then compute the ratio between the asymptotic flux and the flux enclosed within a $r=36^{\prime\prime}$ aperture centered on the CALIFA pointing to estimate the aperture correction factor. We use the deepest survey available to compute the aperture correction factors. Figure \ref{fig:fig2} shows two examples of the asymptotic NUV calculations.

We find that several sources with faint NUV detections (UGC\,03107 and UGC\,09937) are generally sensitive to the maximum radius of the apertures, but visual inspection shows that most NUV emission associated with the host galaxy is confined within the CALIFA aperture, so we retain their aperture correction factor to $1\times$. We find two sources to result in an aperture correction factor $\lesssim 1\times$ (NGC\,1590 and\,UGC 05111), which we set $1\times$. In these cases, the aperture correction factors are within $10\%$ from $1\times$. For NGC\,0675, we find an aperture magnitude significantly larger than the cataloged asymptotic magnitude by \cite{2015yCat..22200006B}. This is likely due to poor background subtraction given the large nearby elliptical galaxy contaminating the background emission. Visual inspection shows that the extent of the UV emission in the GALEX medium imaging survey (MIS) cutout down to the noise level lies entirely within the $36^{\prime\prime}$ aperture, so we take the aperture correction factor to be 1. 

We find that 17 SFGs are not covered in the GALEX footprint. Only MCG+09-22-53 appears to have an optical angular extent smaller than the CALIFA aperture based on the SDSS isophotal diameter of $<1^{\prime}$. We treat all other sources as producing a lower bound on the total H$\alpha$ emission.

We estimate a median aperture correction factor of ${\sim}14\%$, with only 4 sources having an aperture correction factor between ${\sim}2$-$3\times$, and one source NGC 2805 having an aperture correction factor ${\sim}8\times$. Our median correction is smaller than the median correction of 40\% found by \citetalias{2015AA...584A..87C}, but this could be due to using the full circular aperture rather than an elliptical aperture; furthermore, we show in Section 4 and 5 that our derived thermal fractions are consistent with those implied by their H$\alpha$ luminosities. Aperture corrections are listed for all CALIFA SFGs in Table \ref{tab:tab1}.

\subsection{Total H$\alpha$ Flux Fitting}
As done in \citetalias{2015AA...584A..87C}, rather than fitting to individual spaxels, where individual signal to noise ratios may be small, we fit to the summed spectrum, in this case being within $r=36^{\prime\prime}$ circular apertures. Before summing the spectral pixels in the $r=36^{\prime\prime}$ aperture, we interpolate over any bad spectral pixels as indicated by the ``BADPIX'' cube using immediately surrounding good spectral pixels, expanding the number of spaxels used for averaging radially if the number of bad spectral pixels used for averaging remains below $50\%$. 

\begin{figure}
    \centering
    \includegraphics[width=\linewidth]{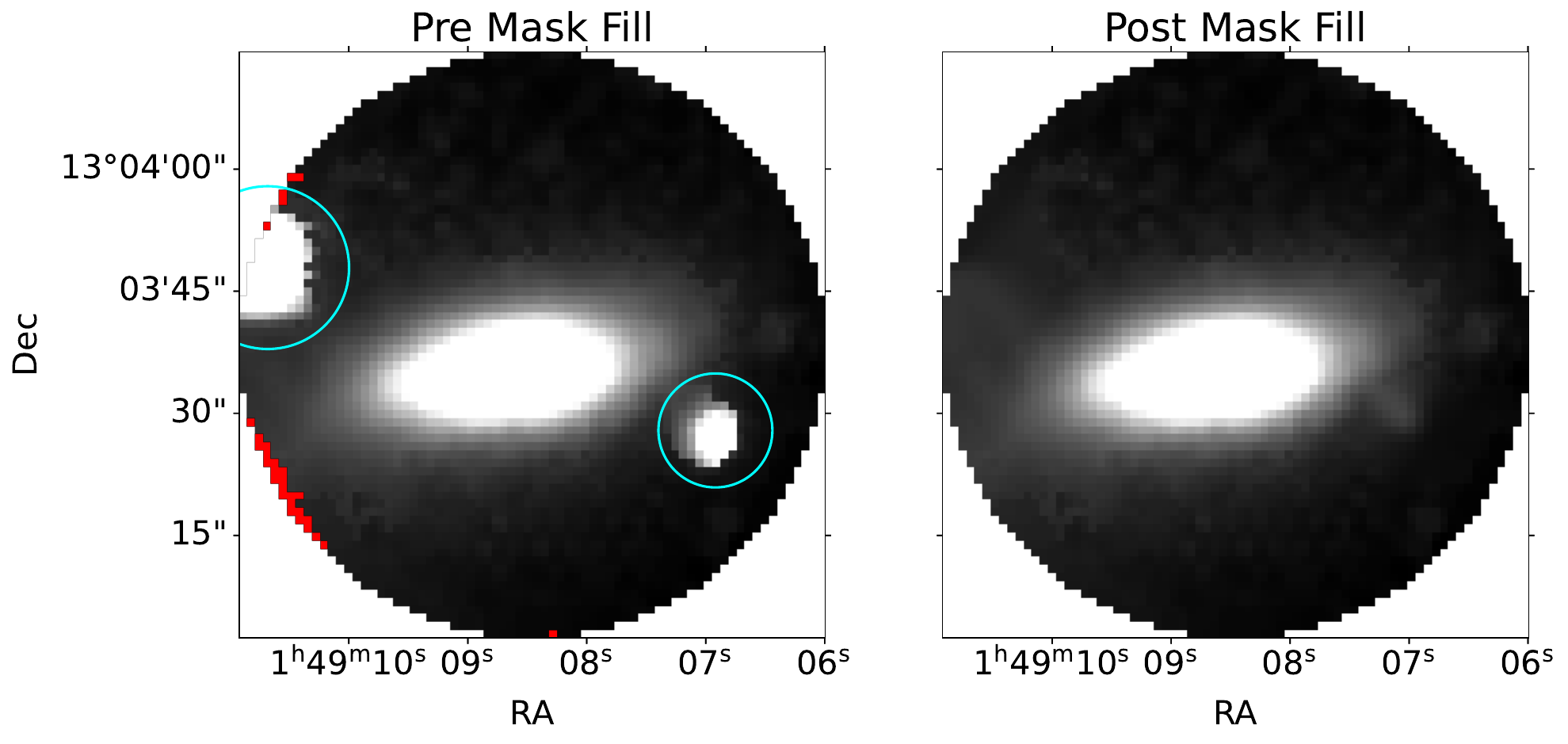}
    \caption{Example of mask-filling of foreground/zero-flux pixels within the $36^{\prime\prime}$ aperture on NGC 0675. Left: cyan apertures indicate the foreground emission that was first interpolated over. The red highlighted regions show the zero-flux pixels within the aperture that were also interpolated over using the non-zero central pixels. Right: the resulting median data cube after interpolation in each spectral pixel.}
    \label{fig:fig3}
\end{figure}

A large fraction of the datacubes have contaminating foreground emission (particularly from non-UV emitting stars) within the $36^{\prime\prime}$ aperture, which are individually identified and masked initially. Furthermore, there are cases where some zero-flux edge pixels remain within the chosen aperture. These masked or empty spatial pixels are interpolated over in each spectral pixel using \textsc{maskfill} (using the same $3\times3$ median filter for iterative interpolation and boxcar filter for final smoothing). In order to examine the effect of interpolating over the foreground/zero-flux spaxels, we fit emission line fluxes over both sets of spectra. For the integrated error spectrum, we simply sum the error contribution from each individual spaxel, as while there is likely covariance between individual spaxels, this is not well understood. In Figure \ref{fig:fig3}, we show an example of a source with and without interpolation over masked regions on the median data cube. We find that for UGC\,10042, significant foreground contamination by a nearby UV-emitting star spread out over $>10^{\prime\prime}$ that is also partially coincident with host galaxy emission makes interpolation unreliable. This is because galaxy-wide emission is smeared into the masked region, causing the interpolated total flux to be significantly higher (${\sim}50\%$) relative to the masked total flux. We instead mask this star and consider the H$\alpha$ luminosity fitted to be a lower bound for this galaxy.

\begin{figure}[htp!]
    \centering
    \includegraphics[width=\linewidth]{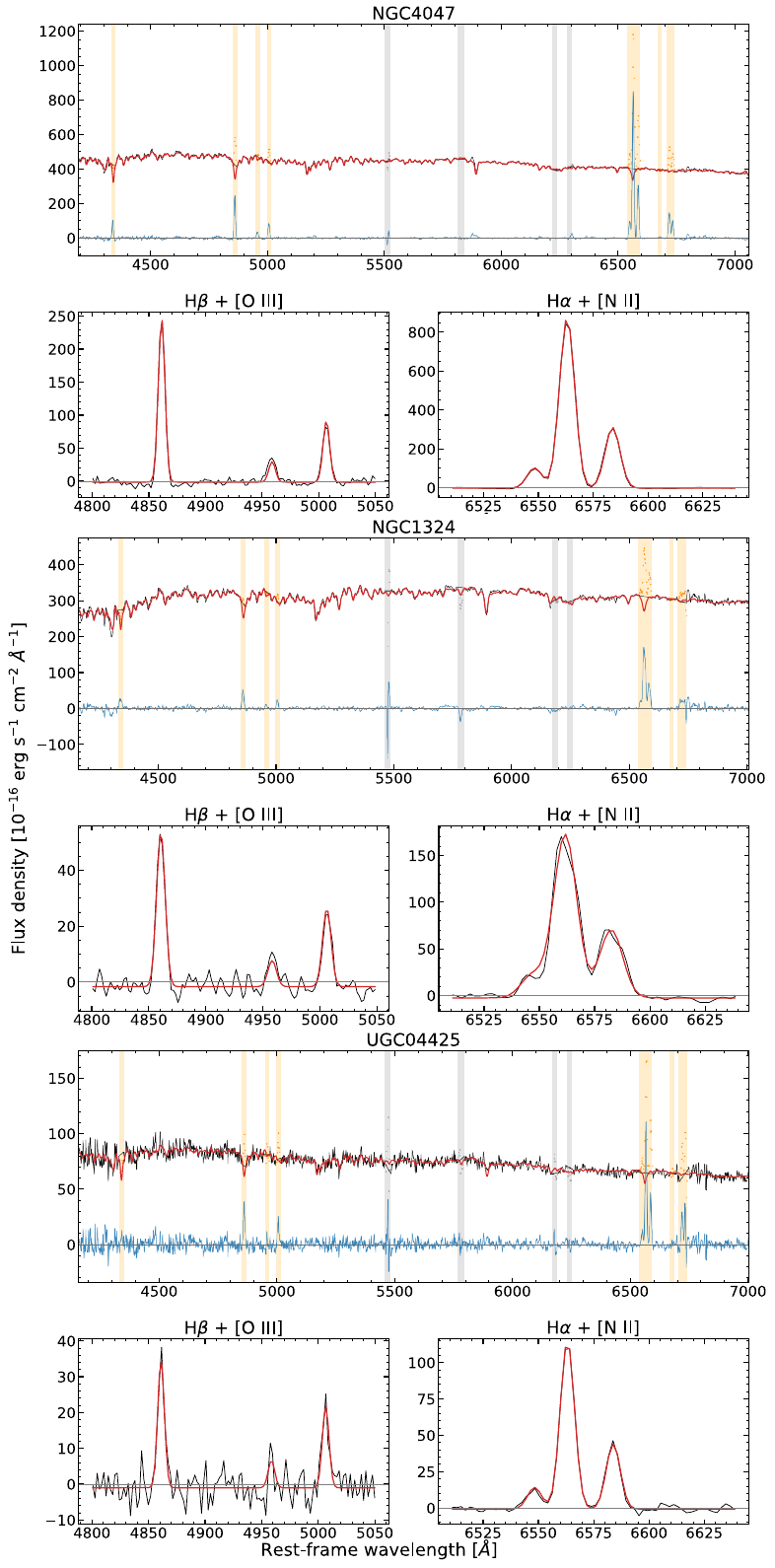}
    \caption{Sample continuum (top panel) and emission line complex (bottom panel) fits for three representative galaxies. NGC 4047 (top) represents a case with strong continuum and nebular features, NGC 1324 (middle) represents a case with strong continuum and weak nebular features, and UGC 04425 represents a case with weak nebular and continuum features. Orange shaded regions represent internally-masked emission line complexes by \textsc{PyPipe3D} intrinsic to star forming galaxies, while gray regions are manually masked locations of sky-lines. Below each continuum fit (red), we show the residual (blue), which represents the nebular spectrum.}
    \label{fig:fig4}
\end{figure}

We adopt a similar approach to the nuclear spectral fitting, using \textsc{PyPipe3D} and the MILES stellar template library to first fit the stellar continuum, masking out typical SFG emission line features and sky-lines. We also only fit over the un-vignetted wavelength range and perform emission line fitting over the same rest-frame wavelength ranges as the nuclear spectral fitting. In Figure \ref{fig:fig4}, we show representative examples of the continuum fits and the corresponding emission line complex fits for galaxies with high signal-to-noise (S/N) nebular, high S/N stellar continuum, and low S/N nebular components, respectively.

The mean correction factor to the raw H$\alpha$ flux due to interpolating over masked foreground stars is ${\sim}1.5$\%, noting that interpolation is used in ${\sim}2/3$ of the cases. We find three sources (UGC\,04240, UGC\,09937, IC\,0307) that, despite having nuclear classifications, do not have H$\beta$ detected to at least $3\sigma$ significance in the \textit{summed} spectrum. This occurs in the case where absorption from the stellar continuum tends to dominate over nebular emission from star forming regions in the summed spectrum or when off-source noisy spaxels in the aperture wash out weak continuum features and thus lead to a poor continuum fit. These sources, however, contain H$\alpha$ detected at the $3\sigma$ level, so we include them as lower bounds on the H$\alpha$ luminosity and hence implied thermal radio emission.

\begin{figure}
    \centering
    \includegraphics[width=\linewidth]{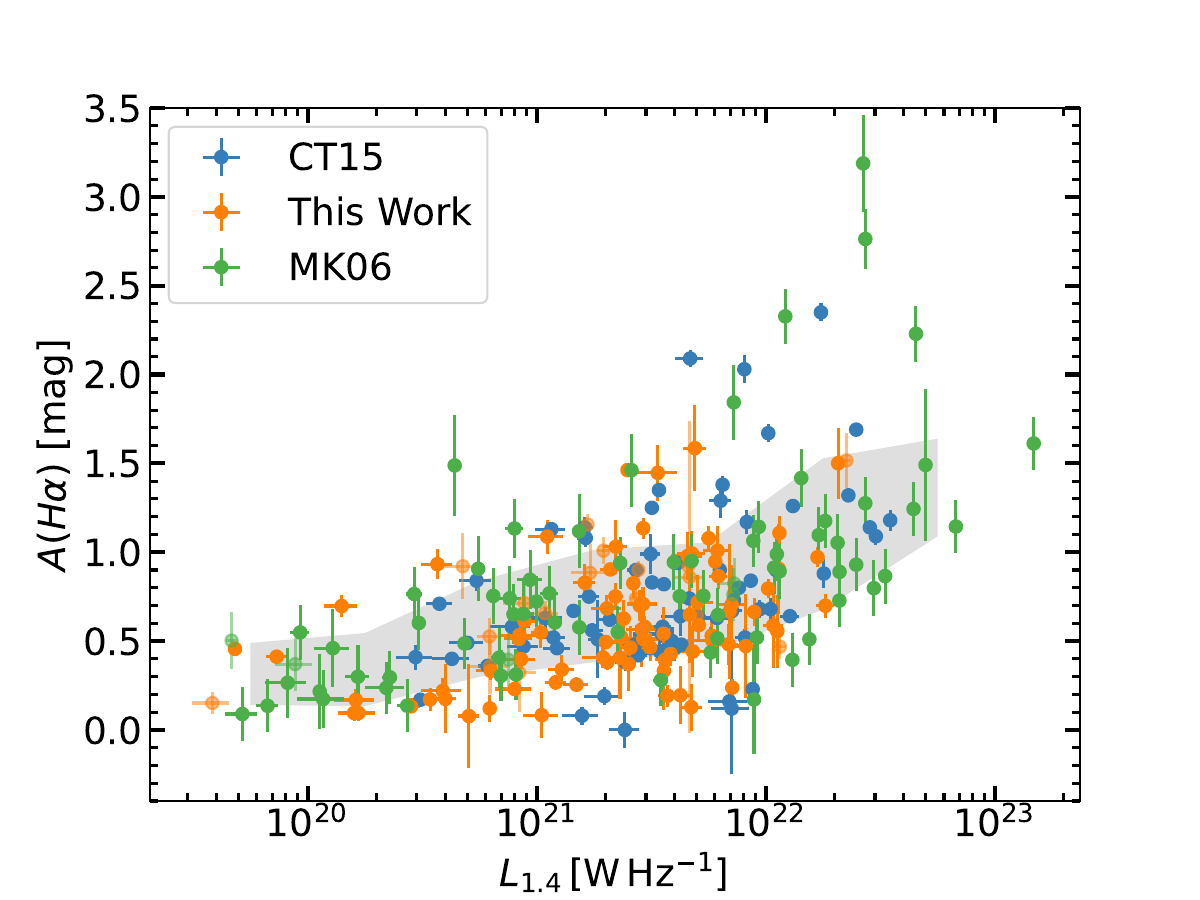}
    \caption{The H$\alpha$ luminosity dust-attenuation as a function of radio luminosity for SFGs considered in \citetalias{2015AA...584A..87C}, this work, and \citetalias{2006ApJS..164...81M}, showing the gradual rise in Balmer decrements with total radio luminosity. An average fixed $\alpha=-0.7$ is assumed for $k$-correction on the radio luminosity; we note that since the maximum $z$ in the sample implied by the observed heliocentric velocities is $z\approx 0.035$, the maximum systematic error incurred on the radio luminosity is $<1.5\%$ for $\alpha<-0.5$, which is negligible. The shaded region represents the ${\sim}1\sigma$ envelope. NGC 0675, which was observed to have $\mathrm{F(H\alpha)/F(H\beta)}<2.86$, is not shown.}
    \label{fig:fig5}
\end{figure}

We follow the same convention as \citetalias{2015AA...584A..87C} to compute the dust-corrected H$\alpha$ luminosities, with
\begin{equation}
    A(\mathrm{H}\alpha)= 5.86 \log\left(\frac{F(\mathrm{H}\alpha)/F(\mathrm{H}\beta)}{2.86}\right)\,\mathrm{mag},
\end{equation}
which is based on the \cite{1989ApJ...345..245C} $R_V=3.1$ extinction law. As auroral lines are not generally available for electron temperature estimation, we assume the theoretical Balmer decrement of $F(\mathrm{H}\alpha)/F(\mathrm{H}\beta)=2.86$ for $n_e=100\,\mathrm{cm^{-3}}$ and $T_e=10^{4}\,\mathrm{K}$. In Section 5.1, we discuss the impact of a fixed electron temperature on the resulting estimated thermal fractions. All sources with the exception of NGC 0675 are consistent to within $1\sigma$ of having $F(\mathrm{H}\alpha)/F(\mathrm{H}\beta)>2.86$, with more radio luminous SFGs generally having larger Balmer decrements (see Figure \ref{fig:fig5}, with sources taken from \citetalias{2015AA...584A..87C} also plotted).

As with the dropped sources, NGC 0675 is mostly continuum dominated with H$\beta$ only detected at a $5\sigma$ level; however, the Balmer decrement is still consistent with negligible dust-attenuation to $2\sigma$. Uncertainties in the total H$\alpha$ luminosities are propagated from the emission line fitting, including the uncertainty in the dust-attenuation. As typical statistical uncertainties in the aperture fluxes and asymptotic magnitudes are $<2\%$ compared to the typical ${\sim}15\%$ uncertainties on the H$\alpha$ aperture luminosity, we ignore their propagation into the total H$\alpha$ uncertainty. Table \ref{tab:tab1} reports the derived dust-attenuation $A(\mathrm{H}\alpha)$ and the resulting total dust and aperture-corrected H$\alpha$ luminosity for our sample of SFGs.

In summary, we are able to estimate total H$\alpha$ luminosities for an additional 103 SFGs (20 being a lower bound) with the CALIFA DR3 MS and ES sample, which are added to the sample of 73 already fitted by \citetalias{2015AA...584A..87C}.

\section{H$\alpha$ Luminosities of the Drift-Scan SFGs}

Similar to the CALIFA dataset, the drift-scan spectra may miss extended star formation associated with a galaxy, which is corrected using GALEX NUV cutouts and the ratio between asymptotic fluxes to the aperture flux associated with the drift scan. For asymptotic magnitudes, we first query \cite{2018ApJS..234...18B} and \cite{2015yCat..22200006B}, computing the aperture magnitude using the image from the same survey, if possible. We found cross matches with 40 sources. If we do not find a cross match, we manually compute asymptotic magnitudes using the same method described in Section 3.1. We were unable to find GALEX cross matches for 5 SFGs in the sample. By inspecting Pan-STARRS1 cutouts \citep{Flewelling_2020}, we find evidence of some extended optical emission beyond the drift-scan coverage, so we treat these remaining 5 sources as producing lower limits on the thermal fractions. For sources with multiple apertures \cite{2006ApJS..164...81M} combine the apertures for a total flux estimate, so we take the union of the apertures before estimating the aperture correction factor for those sources. In this case, because the drift-scan spectra cover an aperture significantly larger than the CALIFA IFUs, we find a median aperture correction factor of just ${\sim}9\%$, with only one source NGC 3628 having $>100\%$. 

H$\alpha$ and H$\beta$ fluxes are obtained by first fitting the continuum using the \cite{2003MNRAS.344.1000B} stellar population synthesis templates including stars with ages from 0 to 12 Gyrs at solar metallicity. In general, there is a degeneracy between age and metallicity when fitting to lower resolution spectroscopy, but the chosen templates are noted to fit the continuum well for most galaxies in the \citetalias{2006ApJS..164...81M} sample. The best fitting stellar continuum spectrum is then subtracted and emission lines are fitted using Gaussian functions. Distances used for luminosity calculations are evaluated in the same manner as those of the CALIFA SFGs. CGCG\,049-057 appears to have an upper limit on the H$\beta$ flux, but that upper limit implies a Balmer decrement $\mathrm{F(H\alpha)/F(H\beta)}>25$, which yields a lower limit to the extinction correction $A(\mathrm{H}\alpha)\gtrsim5\,\mathrm{mag}.$ The implied thermal fraction is at least ${\sim}0.2\%$, significantly lower than other SFGs, so it is likely that most H$\alpha$ and H$\beta$ emission is strongly dust obscured. We therefore drop this SFG from the analysis. We have appended the internal extinctions $A(H\alpha)$ on Figure \ref{fig:fig5}.

In total, we combine H$\alpha$ luminosity measurements of 176 CALIFA, 28 PHANGs, and 74 \citetalias{2006ApJS..164...81M} SFGs.

\section{Results}
For each of the CALIFA, PHANGS, and \citetalias{2006ApJS..164...81M} SFGs, we apply Equation 4, assuming $T_e=10^{4}\,\mathrm{K}$ to predict the thermal radio luminosity. As \cite{2019ApJ...872..148C} do not provide flux densities, we additionally cross match sources to the original NVSS catalog \cite{1998AJ....115.1693C} to retrieve flux densities, with multi-component flux density uncertainties added in quadrature. We find that nine sources do not have matching flux densities between the two catalogs and their constituent components are unknown. Hence, we approximate their flux density uncertainty as $\sigma_f=\sqrt{0.45^2 + (0.03S_{1.4})^2} \,\mathrm{mJy}$, incorporating a $3\%$ uncertainty to the total flux density reflecting uncertainty on the absolute flux density scale \citep[e.g., as done in][]{Matthews_2021a}. These uncertainties may be underestimated for the more extended SFGs, which can span up to ${2}$--$3\times$ the NVSS beam. These uncertainties, along with uncertainties on the H$\alpha$ luminosities are propagated into the thermal fraction calculation for all sources without lower bounds on the thermal fraction. We apply a median $k$-correction $K(z)=(1+z)^{-\alpha-1}$ for $\alpha=-0.7$ for the radio luminosity using the $cz$ heliocentric velocities from \cite{2019ApJ...872..148C}. As noted in Figure \ref{fig:fig5}, the uncertainty on the radio luminosity due to an unknown $k$-correction is negligible (${\sim}1\%$) due to the nearby distances.

\begin{figure*}[t!]
    \centering
    \includegraphics[width=\linewidth]{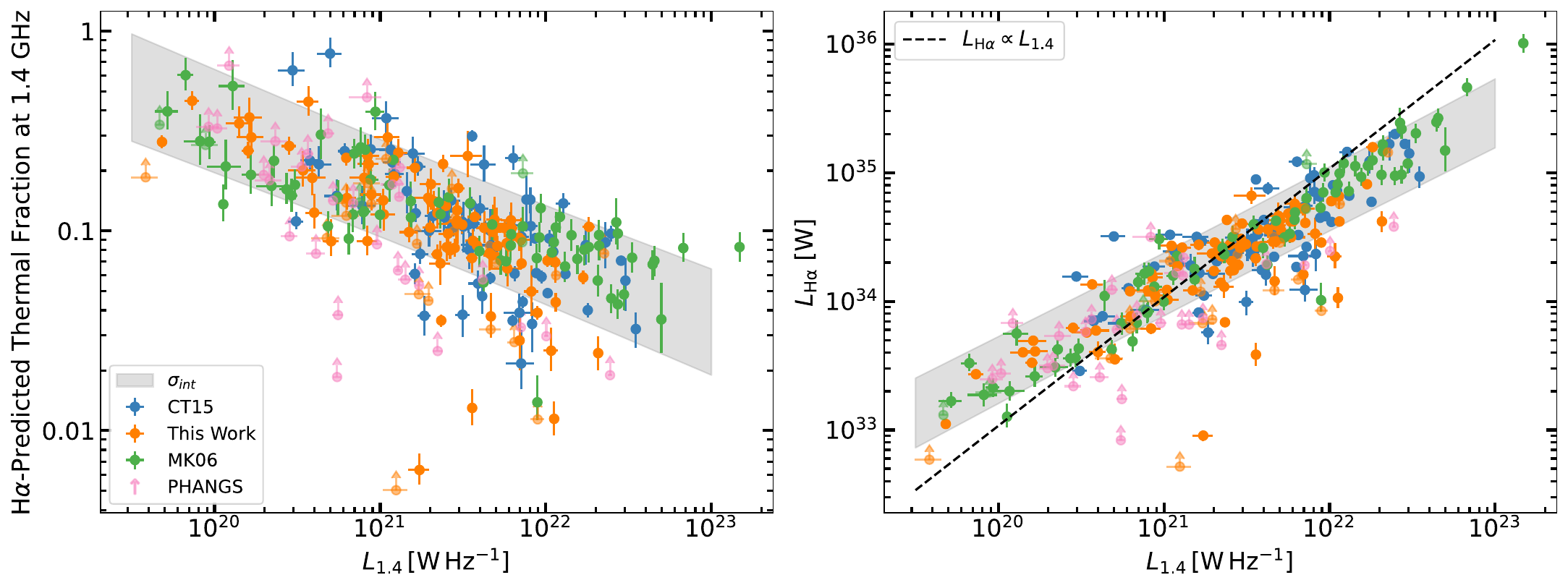}
    \caption{Dependence of the thermal fraction at 1.4 GHz (left) and total dust-corrected, aperture-corrected H$\alpha$ luminosities (right) on the total 1.4 GHz radio luminosity. We have included the implied thermal fractions from total H$\alpha$ luminosities from \citetalias{2015AA...584A..87C}, this work, and \citetalias{2006ApJS..164...81M} and the PHANGS narrow-band H$\alpha$ measurements from \cite{2026A&A...710A.412R}. The shaded region represents the implied intrinsic scatter in the linear relation.}
    \label{fig:fig6}
\end{figure*}

In the case of SFGs where the dust attenuation is unknown (PHANGs), an aperture correction could not be made, or the additional source UGC 10042 where a clean interpolation over the foreground star could not be made, the derived central thermal radio luminosities and fractions are treated as lower bounds. In Figure \ref{fig:fig6}, we show the thermal radio fraction at 1.4 GHz as a function of total radio luminosity. We find that all SFGs are consistent with $f_t<1$ in both the CALIFA and PHANGS samples and that thermal fractions implied by the different H$\alpha$ luminosity fitting methods are generally consistent with each other.

\subsection{Power Law Fits}

We fit a power law to the thermal fractions as a function of total radio luminosity, incorporating intrinsic scatter and uncertainties on both the total radio luminosities and the thermal fractions. This is done using a Markov-Chain Monte Carlo (MCMC) with the package \textsc{emcee} \citep{2013PASP..125..306F}, utilizing a Gaussian log-likelihood and using the survival function for galaxies with lower bounds on the thermal fraction, incorporating the uncertainties on the lower bounds, where available. Given the uncertainties on the radio luminosities that are propagated into the thermal fraction, we also add in quadrature the propagated variance $m^2\sigma^2+2m\sigma^2$, accounting for the covariance between the total radio luminosities and the thermal fractions. Here, $m$ is the power law index and $\sigma$ is the symmetrized $\log(L_{1.4}/(\mathrm{W\, Hz^{-1}}))$ uncertainty. Centered on $L_{\star}\approx 2\times10^{21}\,\mathrm{W\, Hz^{-1}}$ \citep{2019ApJ...872..148C}, we find that $f_t\propto L_{1.4}^{-0.34\pm 0.02}$ with a 1.4 GHz thermal fraction normalization of $f_t(L_{\star})= 0.130\pm 0.005$. We estimate the intrinsic scatter to be $\sigma_{\mathrm{int}}= 0.224\pm 0.012\,\mathrm{dex}$. The resulting linear fit with the intrinsic scatter interval is also shown in Figure \ref{fig:fig6}. 

We find that an independent fit (using a standard Gaussian likelihood) gives $L_{\mathrm{H}\alpha}\propto L_{1.4}^{0.67\pm 0.02}$ with a similar scatter, consistent with the observed scaling for $f_t(L_{1.4})$ to within $1\sigma$, and indicating a sub-linear relationship at the ${\sim} 17\sigma$ level. This relation is somewhat stronger than that found by \cite{2009ApJ...703.1672K, Brown_2017, boselli2018}.

Fitting to just data without lower bounds, we find a slightly flatter but consistent $f_t\propto L_{1.4}^{-0.30\pm 0.02}$. Results are therefore generally robust to the inclusion or exclusion of data treated as lower bounds. 

Ultimately, thermal fractions implied by dust-corrected H$\alpha$ luminosities can change by ${\sim}1\,\mathrm{dex}$ over the observed $L_{1.4}$ range. $L_{\star}/10$ galaxies exhibit thermal fractions that are up to ${\sim}3\times$ larger than $L_{\star}$. Above $L_{\star}$, thermal fractions are similarly observed to decrease by as much as $2$--$3\times$.

\section{Systematics}

\subsection{Assumptions on Electron Temperature}
We first consider the impact of the assumption of a constant average WIM temperature of $T_e=10^{4}\,\mathrm{K}$ due to the non-detection of auroral lines (e.g., [O\,III]$\lambda$4363\,\AA\, or [N\,II]$\lambda$5755\,\AA) in the integrated spectra. This assumption impacts the theoretical Balmer decrement and the dust attenuation correction as well as the ratio of the observed thermal radio emission to H$\alpha$ emission, which has a $T_e^{0.59}$ dependence. More massive SFGs have higher nebular metallicities due to the stronger gravitational potential enabling them retain metals in the presence of galactic winds \citep[e.g., the mass-metallicity relation of SFGs,][]{2004ApJ...613..898T}. The rise in nebular metallicities with SFG luminosity/mass is also supported by the larger observed Balmer decrements in Figure \ref{fig:fig5}. 

Nebular metallicites are also expected to vary inversely with electron temperature \citep[e.g.,][]{2024ApJ...964...47B, 10.1093/mnras/stag149} for both individual HII regions and entire star forming galaxies as larger metal content enables for additional cooling. We estimate nebular metallicites in the sample of SFGs used for H$\alpha$ fitting in this work and the \citetalias{2006ApJS..164...81M} sample using the strong line ratio $\log(\mathrm{O_3N_2})$ calibration from \cite{2017MNRAS.465.1384C} valid for $7.6<12 + \log(\mathrm{O/H}) < 8.85$. We estimate metallicities ranging between $8.3\lesssim 12 + \log(\mathrm{O/H})\lesssim 8.8$, with a preference for more luminous SFGs having higher metallicites, as expected. Averaging in bins of $0.5\,\mathrm{dex}$ in $L_{1.4}$, the average metallicity, however, only changes by ${\sim}0.15\,\mathrm{dex}$, which produces a $\Delta T_e\approx 1500\,\mathrm{K}$ based on the relations provided by \cite{10.1093/mnras/stag149}. At an assumed density of $n=100\,\mathrm{cm^{-3}}$, the theoretical Balmer decrement is expected to vary weakly with temperature $F(\mathrm{H}\alpha)/F(\mathrm{H}\beta)=2.86 T_4^{-0.07}$ \citep{1986A&A...155..297C}. Thus, assuming the dominant effect is the temperature dependence on the thermal fraction $T_e^{0.59}$, for WIM temperatures centered around $10^{4}\,\mathrm{K}$ at $L_{\star}$, we only expect thermal fractions at low (high) radio luminosities to be under(over)-estimated by ${\sim5}\%$ relative to $L_{\star}$ on average due to the fixed temperature assumption. This would slightly steepen the inferred power law index on the thermal fraction-radio luminosity relation.

Additional uncertainties, however, may arise due to assumptions on the average WIM temperature. In particular, the metallicity-temperature relations of \cite{10.1093/mnras/stag149} imply $\langle T_e\rangle\approx 6100\,\mathrm{K}$ for a homogeneous medium compared to $\langle T_e\rangle\approx 7800\,\mathrm{K}$ for an inhomogeneous medium. A reduction in the average temperature by up to ${\sim}40\%$ from assumed can reduce the normalization in the thermal fraction by up to ${\sim}26\%$, but is still broadly consistent with the ${\sim}10\%$ thermal fractions observed at $L_{\star}$.

Finally, the aforementioned metallicity and temperature relations are based on HII regions, while a comparable fraction of H$\alpha$ and thermal radio emission may also come from the less dense DIG. In the Milky Way, WMAP observe a radio thermal free-free to H$\alpha$ emission ratio that implies $T_e\approx 3000\,\mathrm{K}$ \citep{Dong_2011}; this is because some H$\alpha$ emission may be reflected by dust grains, and a large fraction of diffuse radio and H$\alpha$ emission may also come from actively cooling gas \textit{not} in photoionization equilibrium. Equation 4, however, assumes photoionization equilibrium in the production of H$\alpha$ and thermal free-free emission and thus their tie to the same ionizing photon rate. If these observed conditions reflect conditions of the DIG in other SFGs, then assuming that ${\sim}$50\% of H$\alpha$ emission comes from the DIG, thermal fractions would be over-predicted by up to ${\sim}35\%$. This is roughly equivalent to reducing the temperature normalization to $T_e{\sim}6000\,\mathrm{K}$. Under this temperature assumption, we find an $L_{\star}$ thermal fraction of ${\sim}9.6\%$, still consistent with observations. \cite{2007ApJ...661..801O} show that the DIG fraction does not correlate significantly with total H$\alpha$ luminosity and hence total radio luminosity; therefore, we do not expect this systematic to affect the slope of the observed thermal fraction relation significantly. The DIG fraction does, however, correlate strongly with the star formation surface density.

In summary, the systematic effects arising from assumptions on temperature and the separation of DIG and HII region conditions do not appear to significantly affect the power law-slope of the dependence of the thermal fraction. Instead, they tend to affect the overall thermal fraction normalization, which at $L_{\star}$ would still be consistent with the ${\sim}10\%$ thermal fractions that are observed.

\begin{figure*}[t!]
    \centering
    \includegraphics[width=\linewidth]{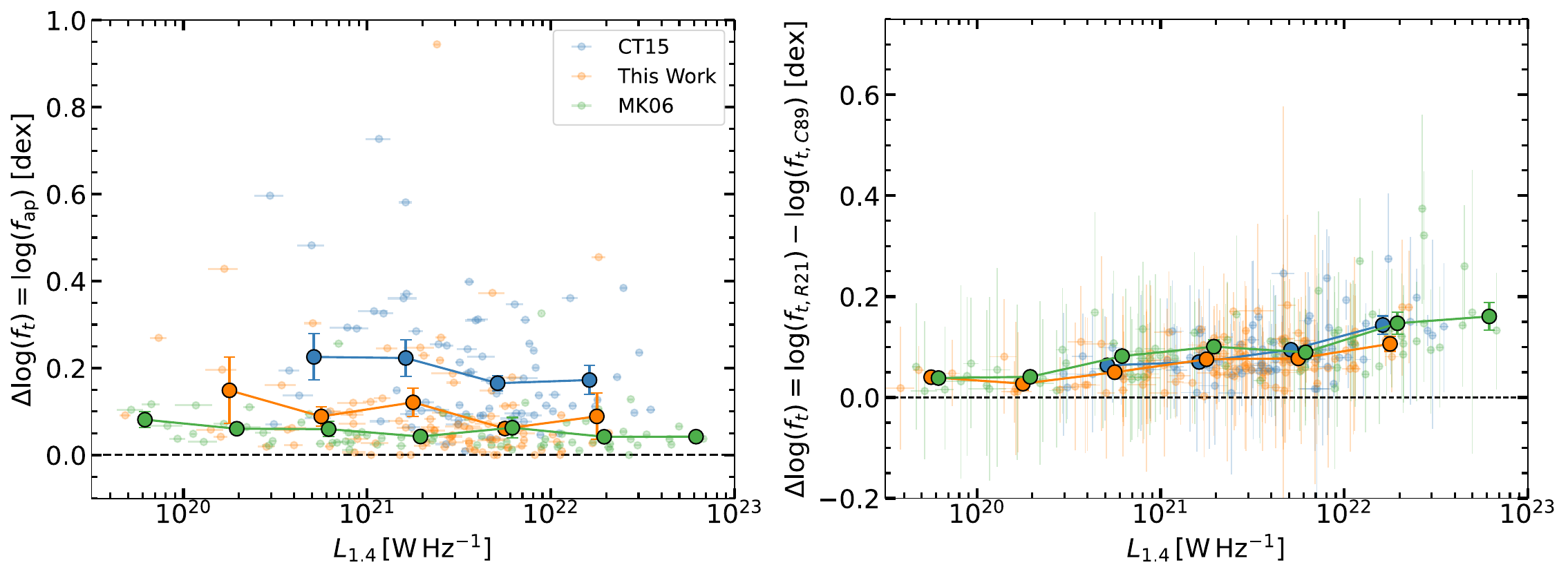}
    \caption{Left: There is only a weak relationship between radio luminosity and the aperture correction factor contribution to $\log(f_t)$. Right: A change to the dust attenuation curve from \cite{1989ApJ...345..245C} (C89) to \cite{10.1093/mnras/stab1885} (R21) increases thermal fractions more at the highest radio luminosities which exhibit the largest Balmer decrements. This yields thermal fractions that are higher by $\lesssim 0.2\,\mathrm{dex}$ on average across all radio luminosities considered. Large solid lines indicate the bin-averaged $\Delta \log(f_t)$ in $\Delta\log(L_{1.4})=0.5\,\mathrm{dex}$ bins.}
    \label{fig:6b}
\end{figure*}

\subsection{Aperture Corrections and Fitting Procedures}

We also consider the overall impact of aperture corrections and the generally different procedures used by \citetalias{2015AA...584A..87C}, this work, and \citetalias{2006ApJS..164...81M}. We consider fitting the observed thermal fraction relation using the SFG samples from the different works. We find power law indices $m=-0.38\pm 0.06, -0.39\pm 0.05, -0.25\pm 0.02$ for the three datasets, respectively. The CALIFA datasets are generally consistent with each other and the combined fit, but the \citetalias{2006ApJS..164...81M} has a weaker dependence relative to the combined fit at a ${\sim}3\sigma$ level and a significantly tighter scatter ${\sim}0.16\,\mathrm{dex}$ in the power law fit. This dependence is still ${\sim}11\sigma$ significant relative to a flat thermal fraction with radio luminosity. This difference may be due to a combination of systematics introduced by (i) the larger CALIFA aperture corrections generating more scatter in $f_t$, (ii) the coarser spectral resolution and different stellar spectral templates, which may introduce systematic uncertainties in the emission line fitting relative to the CALIFA dataset, or (iii), there being significantly more SFGs with $L_{1.4}>10^{22}\,\mathrm{W\, Hz^{-1}}$ relative to the CALIFA datasets, which flatten the power law index.

In the left panel of Figure \ref{fig:6b}, we show the distribution in the $\Delta\log(f_t) = \log(f_{\mathrm{ap}})$ due to NUV-based aperture corrections for SFGs that have them across the three datasets where aperture corrections were performed. As there is only a weak trend on the aperture correction factor with total radio luminosity, the increasing thermal fraction trend still persists with smaller total radio luminosities, with $\Delta\log(f_t)\approx 1\,\mathrm{dex}$ across the full luminosity range. The thermal fractions derived on the SFG sample of \citetalias{2015AA...584A..87C} before aperture correction are lower on average (e.g., near $L_{\star}$) than from this work and \citetalias{2006ApJS..164...81M}, but as discussed in Section 3, this is likely due to using smaller elliptical apertures rather than the full circular aperture coincident with the CALIFA FOV. Indeed, the median percent change in the thermal fractions for the SFG sample in \citetalias{2015AA...584A..87C} is ${\sim}37\%$ compared to ${\sim}14\%$ for our SFG sample. Despite nearby SFGs having a larger angular size and preferentially having lower radio luminosities, the largest aperture correction factors occur at intermediate luminosities near $L_{\star}$, likely because they just dominate the SFG population in flux and volume limited samples. Such SFGs motivate using IFU mosaics or larger FOV IFUs \citep[e.g., SITELLE,][]{2012SPIE.8446E..0UG} in order to rely less on an empirically derived aperture correction.

\subsection{Dust Attenuation Law}

\citetalias{2015AA...584A..87C} note that the choice of the dust extinction curve of \cite{1989ApJ...345..245C} and the dust attenuation curve of \cite{2000ApJ...533..682C} results in just a ${\sim}3\%$ difference in $A(\mathrm{H\alpha})$ at a fixed Balmer decrement. However, more recent dust-law determinations from fits to Balmer lines from stacks of ${\sim}$80,000 local $0.04\lesssim z\lesssim 0.1$ SDSS SFGs indicate that a quadratic fit to the attenuation curve $k(\lambda)$ is preferred over a linear fit \citep[e.g.,][and references therein]{10.1093/mnras/stab1885}. The authors assume a similar electron temperature and density and hence a similar theoretical Balmer decrement as us. In this case, the authors infer a selective absorption $R_V\approx 2.84$ with the pre-factor in Equation 5 increasing to $7.57$. This will preferentially increase H$\alpha$ luminosities when the $\mathrm{H\alpha}/\mathrm{H\beta}$ ratio is larger, particularly at higher radio luminosities. In the right panel of Figure \ref{fig:6b}, we show the change in the inferred thermal radio fraction assuming the dust attenuation curve derived from \cite{10.1093/mnras/stab1885}, which generally increases thermal fractions up to an average of ${\sim}0.2\,\mathrm{dex}$ by the highest radio luminosities. Correspondingly, we find a somewhat weaker dependence $f_t\propto L_{1.4}^{-0.28\pm 0.03}$ with a larger thermal fraction $f_t(L_{\star})=0.144\pm 0.006$. The intrinsic scatter is also slightly larger, with $\sigma_{\mathrm{int}}=0.251\pm 0.013$.

We conclude that assuming a more recent quadratic dust attenuation curve does not break the strong radio-luminosity dependence on the H$\alpha$-inferred thermal fraction but helps increase thermal radio fractions at higher radio luminosities where the dust attenuation is larger.

\subsection{Spectral Signal-to-Noise Ratio and SFG Distance}
\begin{figure}[t!]
    \centering
    \includegraphics[width=\linewidth]{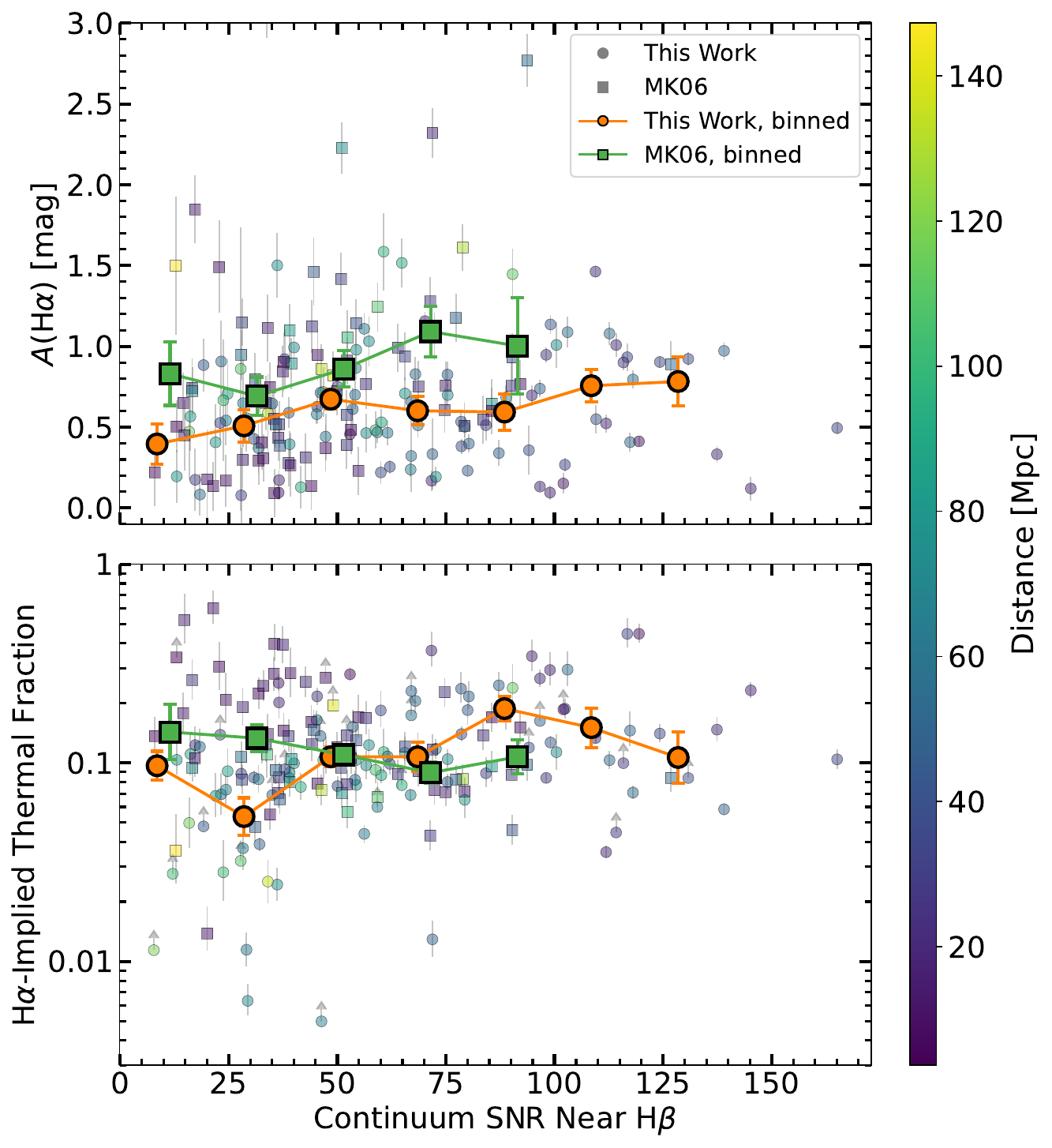}
    \caption{Top: H$\alpha$ extinction increases weakly with the continuum SNR near H$\beta$ across both the datasets in this work and \citetalias{2006ApJS..164...81M}. Bottom: there is no strong correlation with the resulting thermal fraction $f_t$. Binned extinctions and thermal fractions are calculated in intervals of 20 in SNR. Thermal fraction means are done in log-space and only use non-lower-limit values, for simplicity.}
    \label{fig:6.75}
\end{figure}

We investigate whether there are any correlations between the spectral signal-to-noise ratio (SNR) and the obtained thermal fraction and Balmer decrements. As SNRs may be overestimated from the summed CALIFA datacube errors alone (due to spaxel covariances), we instead use an empirical estimator for the SNR in an emission-line free region near the H$\beta$ line using the median absolute difference between adjacent spectral pixels \citep[e.g.,][]{2008ASPC..394..505S}. These SNRs nonetheless correlate tightly with the median signal-to-noise calculated from the integrated error spectrum. SNRs are computed only for the SFGs analyzed in this work and from \citetalias{2006ApJS..164...81M} as we have access to their integrated spectrum. 

We find a weak increasing $A(\mathrm{H}\alpha)$ (or equivalently, the Balmer decrement) with continuum SNR and no correlation with the H$\alpha$-implied thermal fractions. An explanation for the former effect is unclear but may be related to the possibility that lower spectral SNR disfavors detections of obscured H$\beta$ emission, but a selection based on spectral SNR was not explicitly made in the datasets. Given the large scatter, the weak increase in $A(\mathrm{H}\alpha)$ should be interpreted with caution. For the same continuum SNR, the \citetalias{2006ApJS..164...81M} appear to produce somewhat larger Balmer decrements, which may reflect differences in spectral fitting methods, stellar templates used, and spectral resolution. The latter effect suggests that the SNR near H$\beta$ does not strongly affect the luminosity-thermal fraction relation ultimately, with average thermal fractions hovering near the mean thermal fraction of the population which is ${\sim}10\%$.

We do note, however, that the closest SFGs tend to have larger thermal fractions and smaller Balmer decrements while the more distant SFGs tend to have lower thermal fractions and larger Balmer decrements. This is largely a reflection of the intrinsic radio luminosity correlation and radio luminosity selection effect. Lower radio luminosity SFGs are only detected at smaller distances; SFGs with $L_{1.4}=10^{20}\,\mathrm{W\, Hz^{-1}}$ can only be detected out to ${\sim}20\,\mathrm{Mpc}$ given the NVSS $5\sigma$ limit of $S_{1.4}\approx 2.5\,\mathrm{mJy/beam}$. Alternatively, given the strong cutoff in the radio luminosity function above $L_{\star}\approx 2\times10^{21}\,\mathrm{W\, Hz^{-1}}$ \citep{2019ApJ...872..148C}, detections of the higher radio luminosity SFGs require a larger volume, and may be more represented at larger distances.

\subsection{Galaxy Inclination}
\begin{figure}
    \centering
    \includegraphics[width=\linewidth]{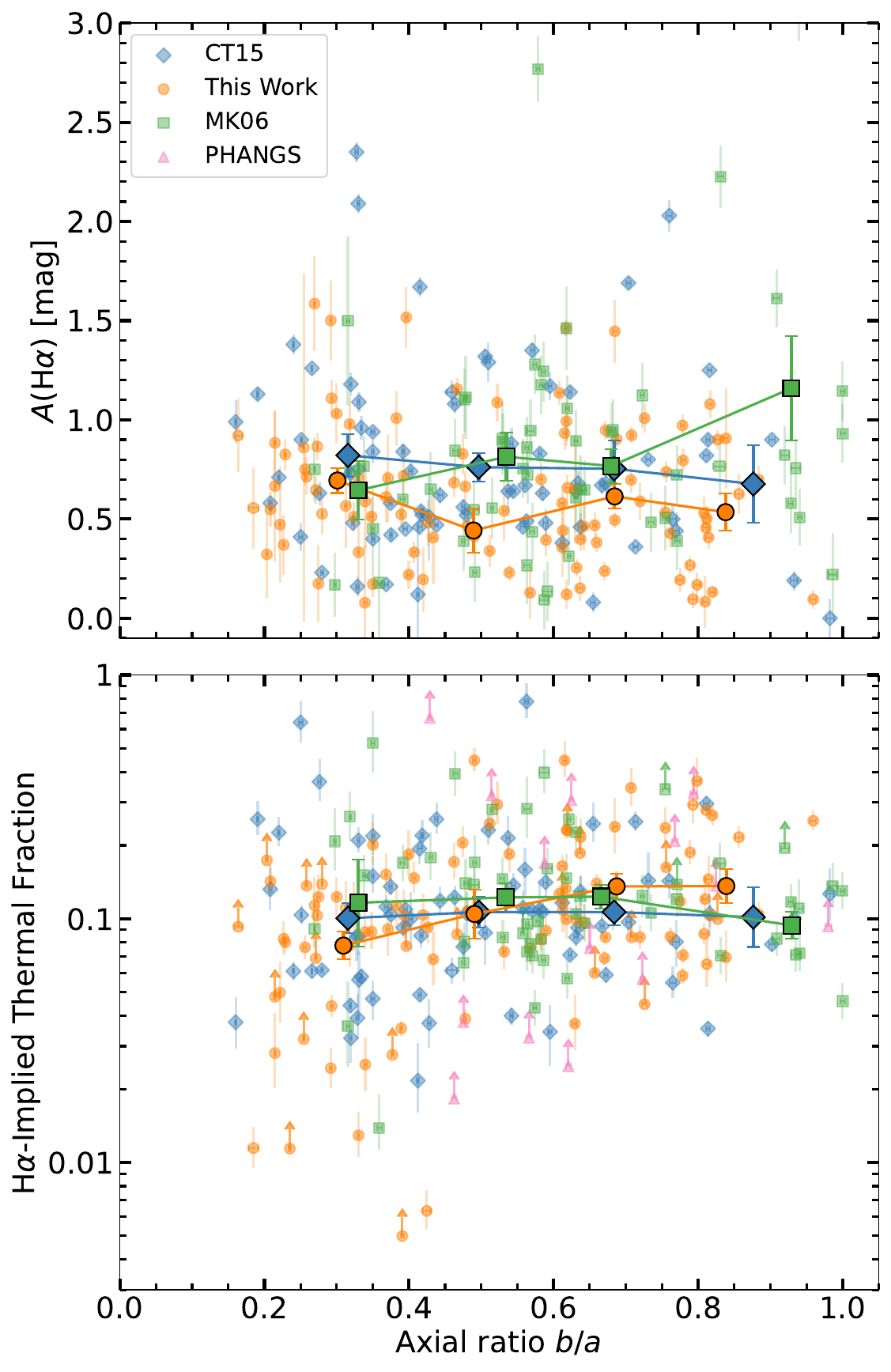}
    \caption{H$\alpha$ extinction (top) and thermal fraction (bottom) as a function of the SDSS exponential fit $b/a$ axial ratio in the $r$-band, if available, which is a proxy for disk inclination. Both quantities show weak correlations with inclination across the entire sample and when averaged in bins of $\Delta(b/a)=0.2.$}
    \label{fig:fig6.9}
\end{figure}

Previous studies have shown that the Balmer decrement is largely independent of the galaxy inclination \citep[e.g.,][]{2010ApJ...709..780Y} despite the possibility that lines of sight through highly inclined galaxies may pass through a larger dust column. To verify this, we query SDSS DR20 for the exponential profile fit minor-to-major axis ratios ($b/a$) in the $r$-band as a proxy for galaxy inclination. As shown in Figure \ref{fig:fig6.9}, we find no strong correlation of the inclination on the axial ratio, with more inclined SFGs (in the lowest $b/a$ bin) having either an $\mathrm{A(H\alpha)}$ or $f_t$ that is not statistically significantly different from less inclined SFGs (in the highest $b/a$ bins) in any sample. Similar to the spectral SNR, the thermal fractions remain on average around the population mean.

We conclude that galaxy inclination is also not a significant driver of the thermal fraction dependence on total radio luminosity derived from H$\alpha$.

\subsection{Comparison to Direct Estimates from Radio SEDs}
As with H$\alpha$-based estimates of the thermal fraction, direct estimates from radio SEDs are not without difficulties. These estimates require observations covering the strongly non-thermal regime at frequencies $\nu \lesssim  1\,\mathrm{GHz}$ and the more thermally dominated regime at frequencies $\nu \gtrsim 33\,\mathrm{GHz}$. This produces a change in $\gtrsim 30\times$ in diffraction-limited angular resolution with the same radio telescope; therefore accurate radio SEDs must generally account for differences in angular resolution. 
\begin{figure}
    \centering
    \includegraphics[width=\linewidth]{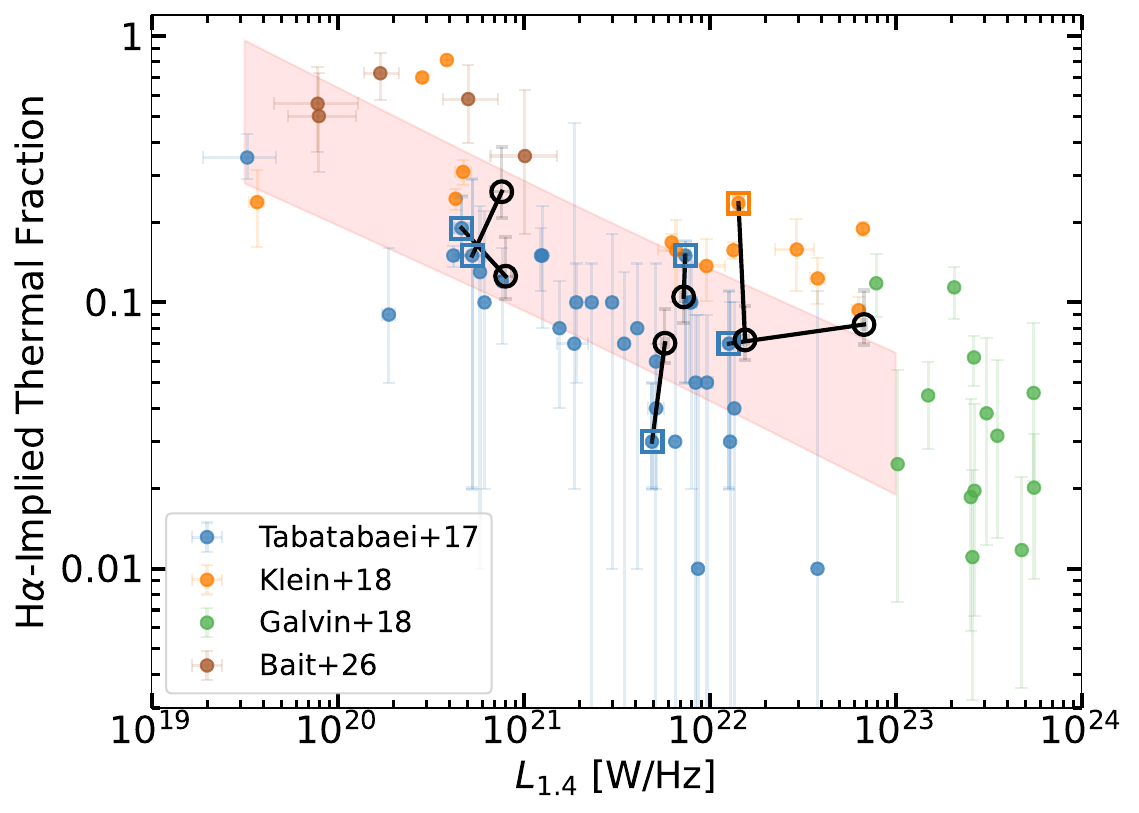}
    \caption{Comparison of the H$\alpha$-derived thermal fractions with radio SED-derived thermal fractions evaluated at 1.4 GHz from \cite{2017ApJ...836..185T, 2018A&A...611A..55K, 10.1093/mnras/stx2613, 10.1093/mnras/stag492}. Overlaid is the fitted power-law with the intrinsic scatter interval (red, shaded). The normalization at high radio luminosities and near $L_{\star}$ is generally consistent with our measurements and show evidence of increasing thermal fractions with lower radio luminosities despite the lack of such SFGs represented in their samples. Solid black lines between colored boxes (literature) and black circles (this work) indicate direct comparisons between radio SED and H$\alpha$-based measurements.}
    \label{fig:fig7}
\end{figure}

Nevertheless, some attempts have been made to constrain thermal fractions directly from radio SEDs in local SFGs, which we can use for comparison with those derived from total H$\alpha$ measurements. Among local SFGs over the luminosity range considered in this work, recent analyses have been conducted by \cite{2017ApJ...836..185T, 2018A&A...611A..55K, 10.1093/mnras/stag492}. \cite{2017ApJ...836..185T} use flux density measurements from $1$ to $10\,\mathrm{GHz}$ to constrain thermal fractions fitting a joint synchrotron and free-free model with constant power law indices. Alternatively, \cite{2018A&A...611A..55K} estimate thermal fractions using SEDs constructed from 300 MHz out to at least 10 GHz and up to 32 GHz. The authors explore fitting radio SEDs with the possibility of the synchrotron component exhibiting a break or cut-off at high frequencies due to cosmic ray electron losses, indicating that even direct estimations from radio SEDs can be strongly model dependent. \cite{10.1093/mnras/stag492} observe metal-poor starbursts at low radio luminosities spanning redshifts $0.02\lesssim z\lesssim 0.08$ and thus not detectable normally with the NVSS. The authors consider models with a pure double power law and additionally with an optically thick component, constraining thermal fractions at 1 GHz, which we use the provided model parameters to estimate at 1.4 GHz (without the H$\beta$-informed Gaussian prior on the 1 GHz thermal flux). Thermal fractions with and without the H$\beta$ prior on the 1 GHz thermal flux are found to be generally consistent with each other, however. In Figure \ref{fig:fig7}, we overlay the direct SED measurements over our data. Despite the different samples being highly inhomogeneous, they exhibit a consistent trend of higher thermal fractions with lower total radio luminosities.

Six SFGs in the \citetalias{2006ApJS..164...81M} overlap with a radio-SED-based thermal radio fraction measurement from \cite{2017ApJ...836..185T} and \cite{2018A&A...611A..55K}, but there are no overlaps with the CALIFA and PHANGs datasets. Among these six SFGs, we find a mean percent difference in $\log(f_t)$ to be ${\sim}11\%$ with the largest being with respect to the single cross match with \cite{2018A&A...611A..55K} at ${\sim}82\%$, indicating some discrepancy between the two methods on an individual SFG level. Radio luminosities also disagree within a factor of a few due to the different distance, peculiar velocity correction conventions, and radio flux densities used. 

This does not necessarily discredit H$\alpha$-predicted thermal fractions because individual SED-fitting methods may also see significant discrepancies and are thus affected by systematic uncertainties. In particular, between \cite{2017ApJ...836..185T} and \cite{2018A&A...611A..55K}, NGC 2146 and NGC 4631 are in common, with \cite{2017ApJ...836..185T} predicting thermal fractions of $f_t=0.01^{+0.1}_{-0.12}$ and $f_t=0.1^{+0.04}_{-0.05}$  and \cite{2018A&A...611A..55K} predicting thermal fractions $f_t=0.189^{+0.011}_{-0.011}$ and $f_t=0.168^{+0.012}_{-0.012}$ at 1.4 GHz, respectively, indicating some level of discrepancy between methods as well. Additionally, at the high luminosity end $L>L_{\star}$, thermal fractions estimated from \cite{2018A&A...611A..55K} are generally higher than estimated from \cite{2017ApJ...836..185T}, which is most likely due to the model dependence of the SED fitting. \cite{2018A&A...611A..55K} show that when fitted using separate thermal and non-thermal components with a fixed power law, these SFGs are inferred to have significantly lower thermal fractions (in some cases consistent with $f_t=0$). Allowing for spectral curvature (e.g., a non-thermal break or cutoff) can produce a larger thermal fraction at 1.4 GHz, highlighting the model-dependent nature of thermal fraction estimation via radio SEDs. Consequently, in addition to systematics in H$\alpha$ fitting, there are significant systematic uncertainties associated with direct SED fitting that can produce individual discrepancies between H$\alpha$-derived and SED-derived measurements.

We also incorporate direct SED measurements for starburst galaxies with $L_{1.4}> 10^{22.5}\,\mathrm{W\, Hz^{-1}}$ from \cite{10.1093/mnras/stx2613}, which although outside the luminosity coverage of our sample, are not inconsistent with the power law fitted in the lower luminosity range and show large scatter. We use the provided \textit{observed} $L_{1.4}$ and $q_{\mathrm{FIR}}$ as done in the original work, noting that these starbursts are at $z<0.16$. As also noted \cite{10.1093/mnras/stx2613}, the three highest luminosity SFGs $>10^{24}\,\mathrm{W\, Hz^{-1}}$ have $q_{\mathrm{FIR}}$ that are $>2\sigma$ deviant from the average (linear) $q_{\mathrm{FIR}}$ found by \cite{2001ApJ...554..803Y} and below the predicted $q_{\mathrm{FIR}}\approx 2.19$ from \cite{Matthews_2021}, which could indicate additional AGN contribution to the radio emission. Similar to \cite{2018A&A...611A..55K}, \cite{10.1093/mnras/stx2613} explore fitting different models to the SED, including those with free-free absorption.

\cite{2017ApJ...836..185T} show that thermal fractions derived from H$\alpha$ luminosities tend to agree with their radio-SED measurements (despite uncertainties of $>50\%$). Indeed, at higher luminosities, our thermal fractions appear to be more consistent with \cite{2017ApJ...836..185T} than with \cite{2018A&A...611A..55K}.
The lack of coverage of SFGs at $L_{1.4}\lesssim 10^{20}\,\mathrm{W\, Hz^{-1}}$ highlights the necessity of constraining thermal fractions in the low mass/luminosity regime directly, which can only be done with sensitivities deeper than that of the NVSS at a similar resolution.

\section{Origin of the Non-linear radio-FIR Correlation}
We show that the rise in thermal fractions with radio luminosity below $L_{1.4}\approx 10^{22.5}\,\mathrm{W\, Hz^{-1}}$ can explain the non-linear radio-FIR correlation observed below this same luminosity threshold \citep[e.g.,][]{Matthews_2021, 10.1093/mnras/stab746, 10.1093/mnras/staf1536}. These observations show that $q_{\mathrm{FIR}}$ is observed to rise with lower $L_{1.4}$.

\begin{figure*}
    \centering
    \includegraphics[width=\linewidth]{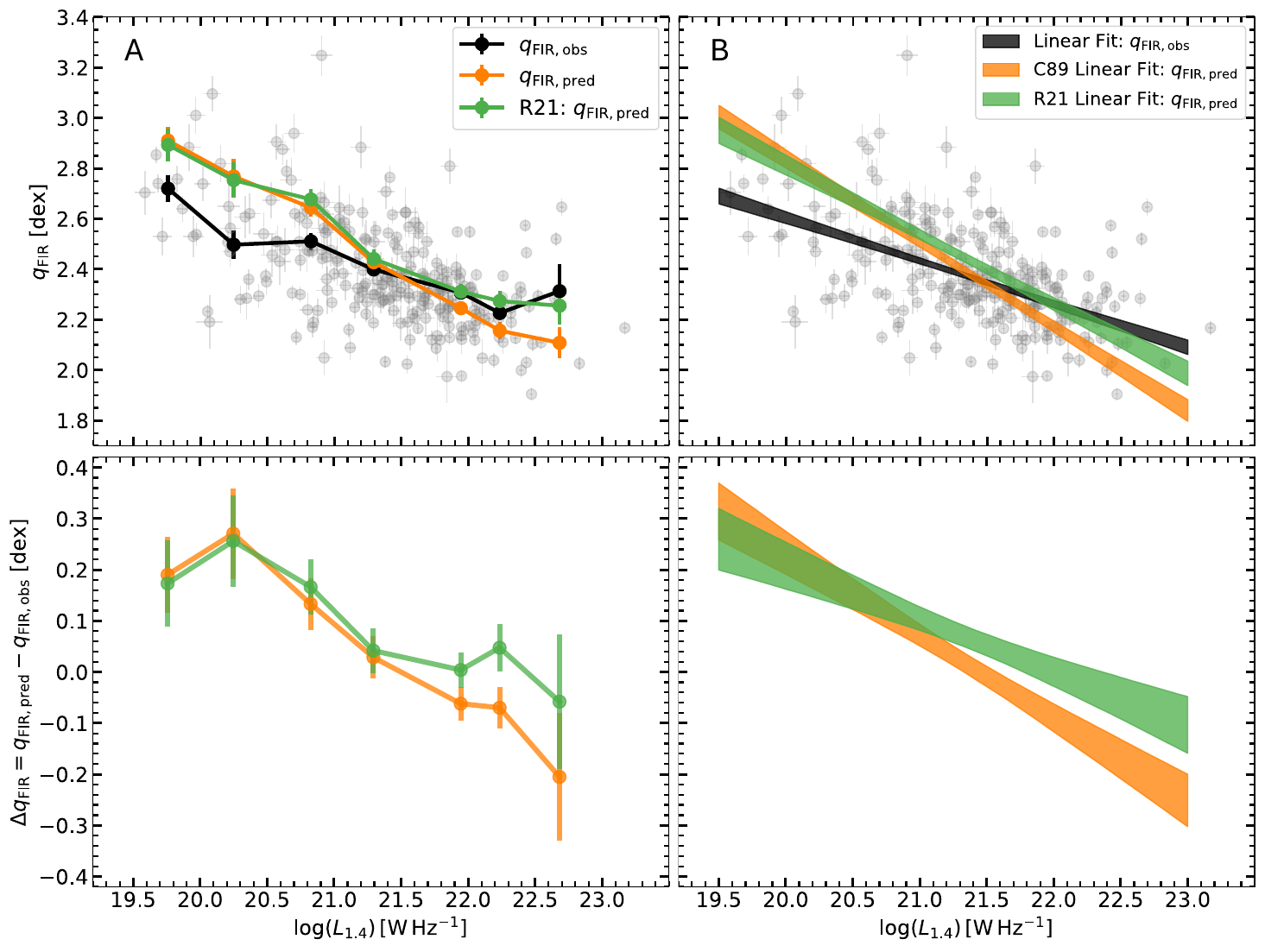}
    \caption{The observed (black) vs. predicted (orange/green) $q_{\mathrm{FIR}}(L_{1.4})$ averaging in $\Delta \log(L_{1.4}/(\mathrm{W\, Hz^{-1}}))=0.5\,\mathrm{dex}$ bins of radio luminosity (left, A) and using the linear fits directly (right, B). The shaded linear regions in panels B represent the $1\sigma$ interval bootstrapped from an MCMC power-law fit with intrinsic scatter. In the top panels, we have overplotted $q_{\mathrm{FIR,obs}}$ (gray) for the 260 SFGs across all the samples with reliable IRAS flux densities. We have considered comparisons with thermal fractions implied by the \cite{1989ApJ...345..245C} (C89, orange) and \cite{10.1093/mnras/stab1885} (R21, green) dust laws in both cases.}
    \label{fig:fig8}
\end{figure*}

From Equations 1, 2, and 3, under the assumption that thermal radio and FIR are perfect star formation tracers and can be calibrated to the same star formation rate quantity according to Equation 4, 10\% thermal fractions in $L_{\star}$ galaxies can reproduce the observed average local $q_{\mathrm{FIR}}\approx 2.34$ from \cite{2001ApJ...554..803Y}. However, as the amount of thermal emission changes with total radio luminosity, Equation 2 predicts a changing $q_{\mathrm{FIR}}$ with radio luminosity. By operating under the \textit{incorrect} assumption that the star formation rate calibration between thermal radio and FIR emission holds across the entire sample as it does near $L_{\star}$, we fix $q_0=3.32$ and estimate the predicted $q_{\mathrm{FIR}}(L_{1.4})$ trend using the observed thermal fractions. Any deviations from the observed $q_{\mathrm{FIR}}(L_{1.4})$ trend indicate deviations from the assumed SFR calibration between FIR emission and thermal radio emission.

For a fair one-to-one comparison, we only use the non-linear radio-FIR correlation derived on the sample of galaxies used in this work. We consider the thermal fractions when derived using the \cite{1989ApJ...345..245C} and \cite{10.1093/mnras/stab1885} dust curves. We define comparison A, predicting $q_{\mathrm{FIR}}(L_{1.4})$ using only SFGs in the full sample that also have robust IRAS FIR flux measurements and averaging $q_{\mathrm{FIR}}$ in bins of $\Delta \log(L_{1.4}/(\mathrm{W\, Hz^{-1}}))=0.5\,\mathrm{dex}$. We only use SFGs where the IRAS quality flag at both $60\,\mathrm{\mu m}$ and $100\,\mathrm{\mu m}$ is $\geq2$ (of which there are 260 out of 278). $L_{\mathrm{FIR}}$ is computed according to the standard IRAS formula used by, for example, \cite{1985ApJ...298L...7H, 1988ApJS...68..151H, 2019ApJ...872..148C}. Because some thermal-fraction measurements are lower bounds, we instead utilize an MCMC, using a Gaussian likelihood for detections and the associated survival probability for lower bounds in each luminosity bin to estimate the bin-mean $\log(f_t)$ and its uncertainties. This requires also fitting for a bin-wise intrinsic scatter. For consistency, this is also done for the $q_{\mathrm{FIR}}$ in each bin. We find intrinsic scatters in $\log(f_t)$ ranging from ${\sim}0.12\,\mathrm{dex}$ in the sparsest bins to ${\sim}0.27\,\mathrm{dex}$ in the most populated bins. Alternatively, we find intrinsic scatters in $q_{\mathrm{FIR}}$ ranging from ${\sim}{0.13}\,\mathrm{dex}$ to ${\sim}0.25\,\mathrm{dex}$ across the bins, but these scatters do not necessarily agree within the same bin.

We also define comparison B using power law fits (with scatter) derived on the observed $q_{\mathrm{FIR}}(L_{1.4})$ trend and that implied by $f_t(L_{1.4})$ fitting only on the IRAS cross-matched SFGs. Our fit is generally consistent with that of \cite{Matthews_2021}, but with a slightly larger slope of $dq/d\log(L_{1.4}/(\mathrm{W\, Hz^{-1}}))= -0.171\pm 0.016$ compared to their $dq/d\log(L_{1.4}/(\mathrm{W\, Hz^{-1}}))\approx -0.147$. Our obtained slope, however, is more consistent with that obtained by \cite{10.1093/mnras/stab746} who find $dq/d\log(L_{1.4}/(\mathrm{W\, Hz^{-1}})) = -0.177\pm0.009$ among their sample.

The resulting comparisons A and B are shown in Figure \ref{fig:fig8}. As observed in the top panels of Comparison A and B, the suppression of synchrotron emission (equivalently, an increase in the thermal fraction $f_t$) drives the non-linear radio-FIR correlation given the generally steeper slope of $\log(f_t)(\log L_{1.4})$ relative to $q_{\mathrm{FIR}}(\log L_{1.4})$. The different slopes, however, also reveal potential inconsistencies in SFR estimation between IR, thermal radio, and dust-corrected H$\alpha$ emission, which can be revealed by taking the difference between the $q_{\mathrm{FIR,obs}}$ and $q_{\mathrm{FIR,p red}}$ curves, as shown in the bottom panels of Comparison A and B.

In particular, we find that $\Delta q_{\mathrm{FIR}} > 0$ going down to lower radio luminosities $L\lesssim L_{\star}$. This is consistent with the analysis of \cite{2003ApJ...586..794B}, indicating that at low radio luminosities, SFGs become optically thin to UV emission, causing IR emission to underestimate the star formation rate. We find that relative to ${\sim}L_{\star}$, IR luminosities may underestimate star formation rates by up to ${\sim}1.9\times$ at ${\sim}L_{\star}/10$ at the $\gtrsim 3\sigma$ level, assuming that thermal radio emission is a perfect star formation tracer near $L_{\star}$. This inferred underestimation is broadly consistent with that inferred by \cite{2003ApJ...586..794B}, who indicate that IR emission may underestimate SFRs by ${\sim}2$--$3\times$ in ${\sim}L_{\star}/100$ SFGs based on IR/UV emission ratios where their $L_{\star}\approx 6.4\times10^{21}\,\mathrm{W\, Hz^{-1}}.$ This is also somewhat consistent with infrared-excess (IR-to-UV ratio, IRX) measurements, where by stellar masses of $M_{\star}\approx 10^{9}\,\mathrm{M_{\odot}}$, $\log(\mathrm{IRX})\approx 0.36$, indicating that IR emission underestimates the IR+UV star formation rate by ${\sim}1.5\times$ \citep[e.g.,][]{10.1093/mnras/stu896}. Stellar masses of $10^{9}\,\mathrm{M_{\odot}}$ are generally consistent with producing radio emission $L_{1.4}\approx 10^{20}\,\mathrm{W\, Hz^{-1}}$ based on the CALIFA sample. 

In comparison A, there is evidence of a possible turnover or flattening in $\Delta q_{\mathrm{FIR}}$ at lowest radio luminosities. Thermal fractions are physically capped at $\log(f_t)=0$ corresponding to $q_{\mathrm{FIR,pred}}=3.32$, so if $q_{\mathrm{FIR,obs}}$ continues to rise with lower radio luminosity beyond our observed range, then $\Delta q_{\mathrm{FIR}}$ must eventually drop. In lower mass/luminosity, metal-poor SFGs, increased Lyman continuum escape may start to reduce the effectiveness of H$\alpha$ emission as a star formation tracer \citep[e.g.,][]{2022A&A...667A..35R}. This is to be interpreted with caution, however, as the sample size in the lowest radio luminosity bin falls to $10$, and this is only a ${\sim} 2.5\sigma$ result. Faint metal poor SFGs as observed in \cite{10.1093/mnras/sty3199}, however, suggest that $q_{\mathrm{FIR}}$ may be lower than normal SFGs at higher radio luminosity, in which case as high thermal fractions are not needed.

Above $L_{\star}$, the differences between the assumed attenuation law become more significant given the larger Balmer decrements observed in this luminosity range. In Comparison A, both $\Delta q_{\mathrm{FIR, obs}}$ and $q_{\mathrm{FIR, pred}}$ appear to flatten, and remain consistent with each other to within $\lesssim 2\sigma$ even in the highest radio luminosity bin. The \cite{10.1093/mnras/stab1885} dust-law produces greater agreement with $q_{\mathrm{FIR,obs}}$ given that $\Delta q_{\mathrm{FIR,obs}}$ is consistent with zero to within $1\sigma$, which would indicate agreement between IR, thermal radio, and dust-corrected H$\alpha$ emission as SFR tracers out to ${\sim}10L_{\star}$. The \cite{1989ApJ...345..245C} dust law, however, seems to indicate a more negative $\Delta q_{\mathrm{FIR}}$ at higher radio luminosities, indicating that some SFGs would exhibit an underestimation of SFR using dust-corrected H$\alpha$ or thermal radio emission relative to IR emission.

The flattening in $q_{\mathrm{FIR, obs}}$ and $q_{\mathrm{FIR,pred}}$ in Comparison A suggests a potential breakdown in the simple power law fit of Comparison B, consistent with flattening in the radio-FIR correlation observed by \cite{Matthews_2021, 10.1093/mnras/staf1536}, but inconsistent with the pure power law fitted by \cite{10.1093/mnras/stab746}. Alternatively, extrapolation of the power law fit to higher radio luminosities seems to suggest a significant underestimation of dust-corrected H$\alpha$ or thermal radio emission relative to IR emission at the $\gtrsim 3\sigma$ level for the \cite{1989ApJ...345..245C} dust curve.

Indeed, at even higher radio luminosities $L_{1.4}>10^{22.5}\,\mathrm{W\, Hz^{-1}}$, $\Delta q_{\mathrm{FIR}}$ may decrease to negative values based on observations. Taking into account the bright starbursts from \cite{10.1093/mnras/stx2613}, for $\langle\log(L_{1.4}/(\mathrm{W\, Hz^{-1}}))\rangle\approx 23.55$ and $\langle q_{\mathrm{FIR}}\rangle\approx 2.26$, and $\langle\log f_t\rangle\approx -1.29$, weighting by the inverse variance. This implies a deficit $\Delta q_{\mathrm{FIR}}\approx -0.23$. Ignoring the three $>10^{24}\,\mathrm{W\, Hz^{-1}}$ which may have AGN contamination, we find $\langle q_{\mathrm{FIR}}\rangle\approx 2.37$, and $\langle\log f_t\rangle\approx -1.16$ implying a deficit $\Delta q_{\mathrm{FIR}}\approx -0.21$, which may suggest a real underestimation in SFRs by thermal radio emission relative to IR emission in luminous starbursts, consistent with the extrapolation of the linear relations in Panel B of Figure \ref{fig:fig8}. A caveat of this is that the $\langle q_{\mathrm{FIR}}\rangle$ among the starburst sample is larger than the $\langle q_{\mathrm{FIR}}\rangle\approx 2.19$ found by \cite{Matthews_2021} in this luminosity regime among their complete sample of SFGs. At such luminosities, Lyman continuum photons could be more significantly absorbed by dust before ionizing neutral hydrogen \citep[e.g., as suggested by SFR comparisons between dust-corrected H$\alpha$ and FIR emission in ][]{2002MNRAS.330..876C}. Alternatively, there may be increasing contributions to FIR emission by old stellar populations \citep{2003ApJ...586..794B, 2011ApJ...737...67M}, but \cite{2003ApJ...586..794B} show that this contamination does not appear to be strongly dependent on IR luminosity and does not affect the radio-FIR correlation. There may also be an increasing AGN contamination in the IR despite being radio quiet in ultra luminous infrared galaxies (ULIRGs) that tend to populate this radio luminosity regime \citep[e.g.,][]{Dey_2024}, thus producing an IR excess relative to radio emission from star formation alone.

We note that in the case where the effective electron temperature normalization is lower by ${\sim}40\%$ (as discussed in Section 6.1) causing the inferred thermal fractions $\log(f_t)$ to drop by ${\sim}0.13\,\mathrm{dex}$ the $q_0\propto \log(T_e^{-0.45})$ dependence implies that $q_0$ will increase by ${\sim}0.1\,\mathrm{dex}$. This results in a net reduction of the inferred $\Delta q_{\mathrm{FIR}}$ by just ${\sim}0.03\,\mathrm{dex}$, which would reduce the significance of the deficit of IR emission relative to thermal radio emission at low radio luminosities down to ${\sim}2.7\sigma$. However, this does not change the conclusion that suppression of synchrotron emission drives the non-linear radio-FIR correlation. Additionally, as lower luminosity SFGs exhibit lower metallicities and thus potentially higher average electron temperatures $T_e$ relative to higher luminosity SFGs, the expected deficit may be a few percent higher relative to assuming a constant electron temperature across the entire SFG population. Both effects, however, may result in a larger deficit of thermal radio emission relative to IR emission in more luminous SFGs. 

This inference also depends on the assumed $L_{\mathrm{TIR}}/L_{\mathrm{FIR}}$ ratio which is used to tie $L_{\mathrm{FIR}}$ directly to a star formation rate. While $L_{\mathrm{FIR}}$ is usually computed as a linear combination of IRAS 60$\,\mathrm{\mu m}$ and 100$\,\mathrm{\mu m}$ flux densities, $L_{\mathrm{TIR}}$ can be obtained via SED fitting \cite{2003ApJ...586..794B} or also by taking a linear combination of IRAS flux densities \citep{10.1093/mnras/staf1536, 1996ARA&A..34..749S}. SFG sample differences and different methods in computing $L_{\mathrm{TIR}}$ can lead to differences in the inferred ratio. While we have assumed an average ratio of ${\sim} 2$, in the case where $L_{\mathrm{TIR}}/L_{\mathrm{FIR}}\approx 2.29$ as found by \cite{10.1093/mnras/staf1536}, $q_0$ will drop by ${\sim}0.06\,\mathrm{dex}$.

The intrinsic scatter on the local radio-IR correlation is inferred to be between ${\sim}0.15\,\mathrm{dex}$ \cite{10.1093/mnras/stab746} to ${\sim}0.26\,\mathrm{dex}$ \cite{2003ApJ...586..794B} depending on whether a non-linear or linear relation is assumed, the specific SFG sample considered, and the extent to which AGN are filtered out from the sample. Our linear fit to the full sample $q_{\mathrm{FIR}}(L_{1.4})$ produces an intrinsic scatter ${\sim}0.17\,\mathrm{dex}$ across the luminosity range. This is smaller than the intrinsic scatter derived for the thermal fractions (or equivalently, the H$\alpha$ luminosities) using the same sample for which we find $\sigma_{\mathrm{int}}\approx 0.23\,\mathrm{dex}$. However, the true intrinsic scatter in the $f_t(L_{1.4})$ relation is likely smaller than derived in this work because of the use of H$\alpha$ emission as a proxy for thermal radio emission. In particular, differences in the true DIG fraction, electron temperature, free-free optical depths, and dust attenuation, which go into Equation 4, will introduce additional scatter on the observed $L_{\mathrm{H\alpha}}(L_{1.4})$ and hence $f_t(L_{1.4})$ relations. So, the derived intrinsic scatter should be interpreted as that fundamentally on $L_{\mathrm{H\alpha}}$ rather than $f_t$. Isolating the true thermal fraction scatter requires an independent estimate of the thermal fraction as a function of total radio luminosity on a comparable sample of SFGs, which is out of the scope of this work. Nevertheless, scatter in the radio-FIR correlation could be interpreted to come from variations in synchrotron suppression at fixed total radio luminosity and dust content (affecting the amount of reprocessed UV emission). As discussed in the following section, variations in the synchrotron suppression, in particular, will arise from variations in ISM conditions (including density, cosmic ray escape scale height, and magnetic field strength).

In summary, the non-linear radio-FIR correlation and its scatter are likely driven by increasing synchrotron suppression, and FIR emission increasingly underestimating star formation rates relative to thermal radio emission in less luminous SFGs. However, for SFGs $\gg L_{\star}$, the relation is difficult to interpret. Comparison A suggests that thermal radio emission likely continues to trace SFR closely with IR emission whereas comparison B and direct luminous starburst thermal fraction measurements suggests that thermal radio emission may underestimate star formation rates relative to IR. More SFGs in the high luminosity regime are needed to confirm this behavior.

\subsection{Origin of Thermal Fraction Change}

Previous works have suggested that less luminous (massive) SFGs are more thermal due to more efficient cosmic ray diffusion (escape) out of their host galaxies due to their weaker magnetic fields \citep[e.g.,][]{1993ApJ...415...93H, 2010ApJ...717....1L, 2011A&A...529A..94C, 2012MNRAS.423L.127R, 2016A&A...593A..77S, 2018ApJS..234...29H}. Alternatively, more luminous (massive) SFGs will have significantly stronger magnetic field strengths, which enhance synchrotron emission and therefore reduce thermal fractions. In this section, we show that this interpretation is broadly consistent with the observed thermal fraction dependence on radio luminosity. Additionally, the observed dependence can be used to provide constraints on relevant cosmic ray and ISM physics across SFGs of different radio luminosities that may not be well constrained.

In Appendix B, we describe a simple steady-state cosmic ray injection model that predicts the dependence of thermal fraction on radio luminosity, incorporating a combination of cosmic ray loss mechanisms, including synchrotron, inverse Compton (via the Cosmic Microwave Background and interstellar radiation fields), escape, ionization, and bremsstrahlung. Here, we do not account for additional secondary cosmic ray production, which likely becomes more important in luminous starbursts \citep{2010ApJ...717....1L}.

\begin{figure}
    \centering
    \includegraphics[width=\linewidth]{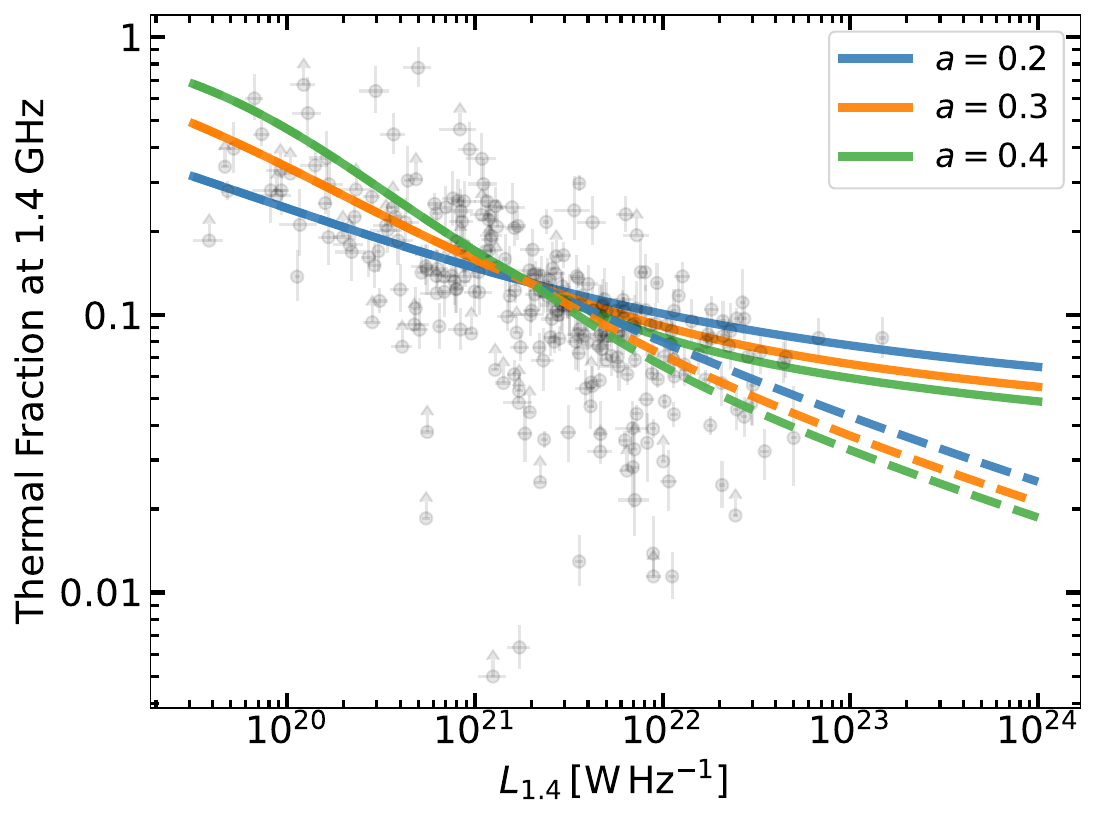}
    \caption{Our fiducial model of cosmic ray electron losses without (solid) and with (dashed) assuming a reduction in the thermal radio emission per unit star formation rate at high radio luminosities as an illustrative example. As expected, thermal fractions remain more suppressed at high radio luminosities and are more consistent with a single linear fit when the thermal radio emission per unit SFR decreases (e.g., as in Figure \ref{fig:fig6} extrapolated out to higher radio luminosities). As the behavior at high luminosities is uncertain, both models are broadly consistent with the observed thermal fractions. The $a=0.3$ scaling represents a magnetic field scaling self-consistent with the observed thermal fraction dependence.}
    \label{fig:fig9}
\end{figure}

Estimating the thermal fraction with the model requires specifying a magnetic field strength prescription to SFGs of different radio luminosities (Equation B6, which is parameterized by a power law dependence $B\propto L_{1.4}^{a}$). In Figure \ref{fig:fig9}, we show the resulting thermal fraction predictions for $a\in[0.2, 0.4]$, which can reasonably replicate the range of observed magnetic fields across the SFG sample, and for the two prescriptions on the thermal radio luminosity per unit star formation rate. We note that $0.2\leq a\leq 0.3$ is generally consistent with the equipartition scaling \citep{2005AN....326..414B}; however, this assumes a fixed thermal fraction, no significant cosmic ray losses, and fixed radio-emitting region size. Additionally, under our observed $L_{\mathrm{H\alpha}}\propto L_{1.4}^{0.67}$, taking dust-corrected H$\alpha$ as the SFR tracer, we have that $\mathrm{SFR}\propto L_{1.4}^{0.67}$. Under our model described in Appendix B, $B\propto \mathrm{SFR}^{0.45}\propto L_{1.4}^{0.3}$, so $a=0.3$ is self-consistent with our observations, which, as shown in Figure \ref{fig:fig9}, models the observed thermal fractions generally well.

We also perform an extrapolation out to higher radio luminosities, noting that SFGs with $\log(L_{1.4}/(\mathrm{W\, Hz^{-1}}))>23$ are more likely to be populated by starbursts with significantly stronger magnetic field strengths, so the prescribed model may not hold. Due to the lack of SFGs with $\log(L_{1.4}/(\mathrm{W\, Hz^{-1}}))>23$, both models can visually replicate the observed thermal fraction dependence, where the primary drive in synchrotron suppression and the scatter in thermal fractions can be attributed to changes in the magnetic field strength at any given radio luminosity.

\begin{figure}
    \centering
    \includegraphics[width=\linewidth]{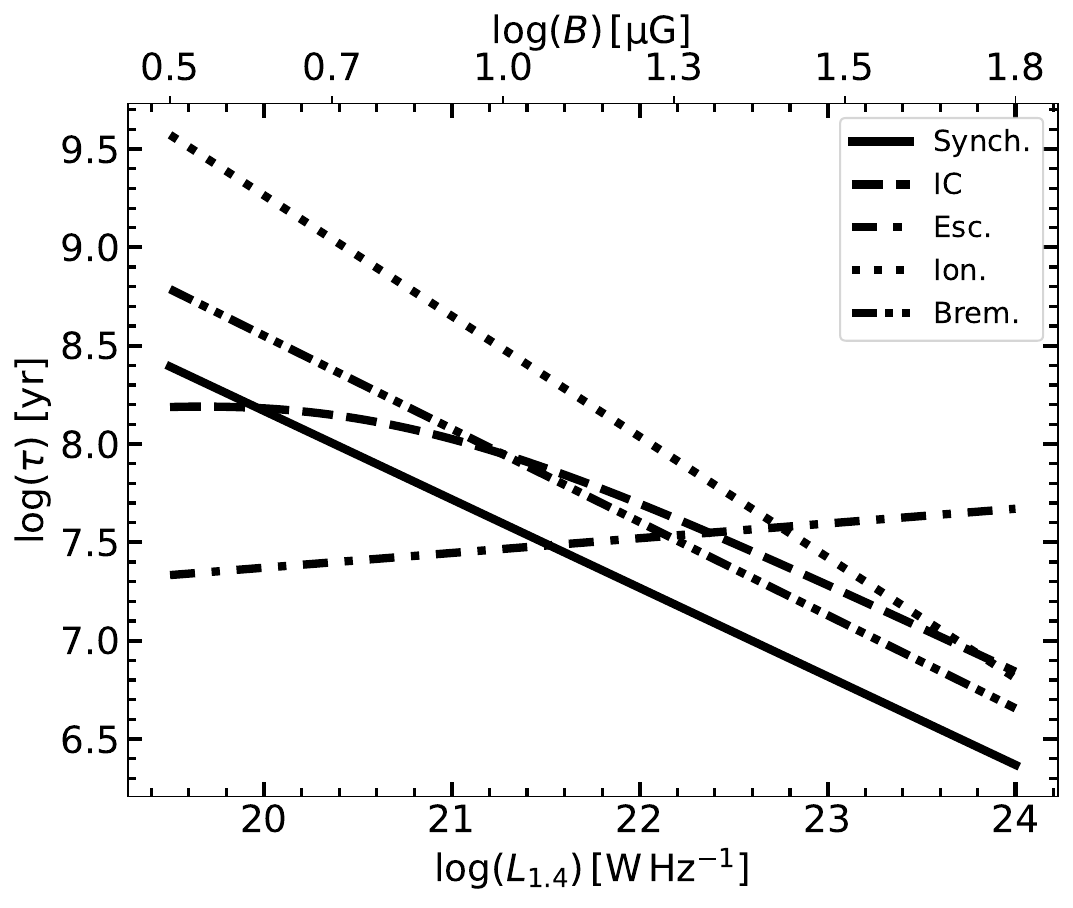}
    \caption{At low radio luminosities, escape losses dominate, while at large radio luminosities, synchrotron losses dominate. Synchrotron suppression in low luminosity SFGs $L<L_{\star}$ can be interpreted as due to cosmic ray escape  within weaker magnetic fields. We have assumed $B{\sim}L_{1.4}^{0.3},$ but these conclusions remain unchanged for alternative magnetic field scalings. This applies for primary cosmic rays.}
    \label{fig:fig10}
\end{figure}

In Figure \ref{fig:fig10}, we show the dominant loss timescales at the cosmic ray electron energy producing $1.4\,\mathrm{GHz}$ emission as a function of the radio luminosity of SFGs. For simplicity, we adopt $a=0.3$ on the magnetic field scaling, generally consistent with the equipartition measurements of \cite{2017ApJ...836..185T}. This results in a range of average magnetic field strengths from $B=3\,\mathrm{\mu G}$ to $B=26\,\mathrm{\mu G}$ over the observed radio luminosity range of our sample and as high as ${\sim}65\,\mathrm{\mu G}$ at $L_{1.4}=10^{24}\,\mathrm{W\, Hz^{-1}}$. The lower end is generally consistent with equipartition measurements of stacked faint dwarf irregular SFGs having $B{\sim}  2\,\mathrm{\mu G}$ and $L_{1.4}\lesssim 10^{19}\,\mathrm{W\, Hz^{-1}}$ \citep[e.g.,][]{2012MNRAS.423L.127R}. The upper end is generally consistent with equipartition magnetic field strengths observed at higher redshifts in the same luminosity range $L_{1.4}\gtrsim 10^{23.5}\,\mathrm{W\, Hz^{-1}}$ \citep[e.g.,][]{Tabatabaei_2025}, inferred to be $B{\sim}55\,\mathrm{\mu G}$ when extrapolated down to $z=0$. However, we note that in the presence of cosmic ray losses, the revised equipartition formula \citep[e.g.,][]{2005AN....326..414B} may not hold. 

Correspondingly, the cause of synchrotron suppression in low luminosity SFGs is consistent with the interpretations of \cite{1993ApJ...415...93H, 2010ApJ...717....1L, 2011A&A...529A..94C, 2012MNRAS.423L.127R, 2016A&A...593A..77S, 2018ApJS..234...29H}. We note, however, that the thermal fraction can also rise purely due to weaker magnetic fields relative to $L_{\star}$ SFGs without requiring a characteristic scale height for cosmic ray diffusion, due to the extra magnetic field strength dependence on $L_{\nu}^{\mathrm{NT}}$ independent of any losses (Equation B4) and reduction in synchrotron power relative to other losses.

At higher radio luminosities, synchrotron losses dominate, but ionization and bremsstrahlung losses are also significant as the ISM density grows with the magnetic field strength, which is responsible for the decline in the rate of decrease in the thermal fractions with radio luminosities above $L_{\star}$ as seen among the solid tracks in Figure \ref{fig:fig9}. These dominating effects in each luminosity regime persist for all magnetic field scalings $a\in [0.1, 1]$ with radio luminosity. The dashed tracks corresponding to a gradual decrease in the thermal radio emission per unit star formation rate can produce lower thermal fractions and explain the discrepancy with FIR emission if thermal fractions continue to drop with higher radio luminosities.

While ionization and bremsstrahlung losses can take away energy of primary cosmic ray electrons, thus slowing the rate of decrease in thermal fractions, the production of secondary cosmic ray electrons from hadronic processes \citep[e.g., pion production;][]{2010ApJ...717....1L} may additionally strengthen synchrotron emission, particularly in dense starbursts. This is because the production rate of secondary cosmic ray electrons will scale with the ISM density and thus be enhanced in more radio luminous SFGs. Consequently, at fixed total radio luminosity, a lower SFR is required, which causes the thermal fraction to drop. However, a lower SFR will also reduce the amount of reprocessed FIR emission, so it is unlikely that secondary cosmic ray electrons can explain the discrepancy between FIR and thermal radio emission at high radio luminosities.

In any case, our fiducial model of cosmic ray electron losses without assuming any additional reduction in thermal emission per unit SFR at high radio luminosity can explain the inverse-variance-weighted thermal fractions of \cite{10.1093/mnras/stx2613} at their average radio luminosity. In this case, an interpretation of excess FIR emission per unit SFR is preferred over a change in thermal or non-thermal emission per unit SFR.

The production of secondary electrons is less significant in less radio luminous SFGs, where the ISM densities are lower. Out to $\lesssim 10L_{\star}$, additional non-thermal contributions by secondary cosmic ray electrons are not needed to explain the observed thermal fractions. However, variations in their population could contribute to the scatter, as thermal fractions well below $10\%$ are observed near $L_{\star}.$

\section{Conclusions}
Using total H$\alpha$ luminosities derived from wide-field integral-field spectroscopy from the CALIFA, narrow-band H$\alpha$ imaging from the PHANGS survey, and drift-scan survey of \citetalias{2006ApJS..164...81M}, we estimate the fraction of radio emission that is thermal at $1.4\,\mathrm{GHz}$ across a sample of $278$ SFGs. We have filtered out possible AGN contamination to construct a clean sample of SFGs using a combination of optical nuclear classifications and infrared emission. In order to estimate total dust-corrected H$\alpha$ luminosities, we perform aperture corrections using observed GALEX NUV emission.

Our primary conclusions are:
\begin{enumerate}
    \item Assuming a pure power law relation between the $1.4\,\mathrm{GHz}$ thermal fraction $f_t$ and the total $1.4\,\mathrm{GHz}$ radio luminosity, we find that $f_t\propto L_{1.4}^{-0.34\pm 0.02}$ with an intrinsic scatter $\sigma_{\mathrm{int}}{\sim}0.22\,\mathrm{dex}$. This results in a change of thermal fractions by ${\sim}1\,\mathrm{dex}$ for radio luminosities between $19.5\leq \log(L_{1.4}/(\mathrm{W\, Hz^{-1}}))\leq 22.75$. We find ${\sim}13\%$ thermal fractions at $L_{\star}\approx 2\times 10^{21}\,\mathrm{W\, Hz^{-1}}.$ The corresponding $L_{\mathrm{H}\alpha}$--$L_{1.4}$ relation is more sub-linear than found previously \citep[e.g.,][]{2009ApJ...703.1672K, 2015A&A...579A.102B, Brown_2017}.
    
    \item Because $\left|d\log(f_t)/d\log(L_{1.4}/(\mathrm{W\, Hz^{-1}})) \right| > \left|dq/d\log(L_{1.4}/(\mathrm{W\, Hz^{-1}}))\right|$, we interpret the non-linear radio-FIR correlation below $\log(L_{1.4}/(\mathrm{W\, Hz^{-1}}))=22.5$ as being driven by increasing synchrotron suppression along with IR emission underestimating star formation rates as total radio luminosities decrease. Independent from appealing to UV-to-IR ratios, we find that IR emission can underestimate star formation rates relative to thermal radio emission (and H$\alpha$ emission) by up to a factor of ${\sim}1.9\times$ by ${\sim}L_{\star}/10$. This underestimation is generally consistent with that found by \cite{2003ApJ...586..794B} who infer an underestimation of ${\sim}2$-$3\times$ by ${\sim}L_{\star}/100$  where their $L_{\star} = 6.4\times10^{21}\,\mathrm{W\, Hz^{-1}}$.

    \item For $L>L_{\star}$, interpretations on the thermal radio-to-FIR calibration are less certain, and motivate further studies using a larger targeted sample in that luminosity range. Thermal fractions averaged in radio luminosity bins suggest that thermal radio and FIR emission both continue to trace star formation well above $L_{\star}$ out to $L_{1.4}=10^{22.5}\,\mathrm{W\, Hz^{-1}}\approx 10L_{\star}$, while the power law fits and direct observations of thermal fractions in starbursts with $L_{1.4}>10^{23}\,\mathrm{W\, Hz^{-1}}$ may suggest that thermal radio emission will eventually underestimate star formation rates relative to FIR emission at high radio luminosities.
    
    \item Intrinsic scatter in the non-linear radio-FIR correlation may be driven by the scatter in the thermal radio emission at 1.4 GHz arising from different ISM conditions (e.g., magnetic field strengths and ISM densities). The observed non-linear radio-FIR correlation in the IRAS FIR cross-matched sample of 260 SFGs exhibits an intrinsic scatter $\sigma_q\approx 0.17\,\mathrm{dex}$. This is below our inferred intrinsic scatter on the H$\alpha$--derived thermal fractions (${\sim}0.23\,\mathrm{dex}$) on the same sample, which is interpreted to be an upper bound.
    
    \item We construct a simple continuous injection cosmic ray loss model incorporating synchrotron, inverse Compton, ionization, bremsstrahlung, and escape losses that is generally successful at explaining the observed $f_t(L_{1.4})$ trend. This model is consistent with the origin of the synchrotron suppression at $L<L_{\star}$ being due to increasing cosmic ray escape losses relative to synchrotron losses, but more fundamentally, weaker magnetic field strengths, corroborating \cite{1993ApJ...415...93H, 2010ApJ...717....1L, 2011A&A...529A..94C, 2012MNRAS.423L.127R, 2016A&A...593A..77S, 2018ApJS..234...29H}.

    \item Thermal fraction reduction at $L\gtrsim L_{\star}$ is driven by synchrotron losses dominating over escape. As ionization/bremsstrahlung losses start to become relevant, however, the rate of decrease in thermal fractions above $\log(L_{1.4}/(\mathrm{W\, Hz^{-1}}))= 22.5$ may slow down. This effect could be counteracted by the decreasing efficiency of thermal radio emission in tracing the star formation rate or enhanced secondary electron production \citep[e.g.,][]{2010ApJ...717....1L}, which are degenerate. The agreement of the fiducial model predictions with the thermal fractions of \cite{10.1093/mnras/stx2613} instead point to an excess of FIR emission per unit SFR.
\end{enumerate}

To investigate the potential underestimation of radio--derived star formation rates at high radio luminosities, direct measurements of thermal radio emission from SEDs \citep[e.g.,][]{10.1093/mnras/stx2613} should be compared to thermal emission implied by total dust--corrected H$\alpha$ luminosities. A deficit arising in both could primarily be attributed to dust absorbing ionizing photons before they are used to ionize neutral hydrogen \citep[e.g.,][]{Inoue_2001, 2003ApJ...583..727D}, but different deficits implied by each method could be more difficult to explain, and interpretations can be complicated by the model-dependence of SED derivations.

Potentially complicating the observed thermal fraction normalization and dependence on total radio luminosity include assumptions on the electron temperature and differing conditions between the WIM within HII regions and within the DIG. Assuming a fixed $T_e=10^{4}\,\mathrm{K}$ across the entire luminosity range and across HII regions and the DIG may introduce systematic errors of up to ${\sim}30$--$35\%$ on the derived thermal fractions.  Additionally, assumptions on the dust attenuation law can have a significant impact on the inferred thermal fractions by on average ${\sim}30\%$, particularly at higher radio luminosities where the Balmer decrements are larger. Nevertheless, the general consistency with thermal fractions derived from direct radio SED constructions indicate that total, dust-corrected H$\alpha$ luminosities are a reliable way to infer typical thermal radio emission over large samples. Next generation radio telescopes such as the ngVLA \citep{2015arXiv151006438C}, SKA \citep{2009IEEEP..97.1482D}, and DSA \citep{2019BAAS...51g.255H}  will be able to provide microjansky sensitivities from $1.4\,\mathrm{GHz}$ up to $90\,\mathrm{GHz}$ and will therefore be important for constraining thermal radio emission from direct SED constructions. This will enable a better evaluation of the effectiveness of dust-corrected total H$\alpha$ emission as a thermal radio luminosity tracer.

To avoid having to do H$\alpha$ luminosity aperture corrections, this work motivates the use of larger FOV IFUs or wide-field narrow-band imaging surveys to better constrain total H$\alpha$ luminosities in nearby SFGs for more direct comparisons to total radio emission. For example, narrow-band imaging with the future Lazuli space observatory \citep{2026ApJ..1010..145R} in both H$\alpha$ and H$\beta$ may provide a way to fold in dust-attenuation corrections directly without having to treat observed narrow-band H$\alpha$ luminosities as lower bounds.

%% Please use the acknowledgment and contribution environments. This will 
%% be anonomyized when the "anonymous" style option is used. 
\begin{acknowledgments}
This work was funded in part by Schmidt Sciences. This work uses data provided by the Calar Alto Legacy Integral Field Area (CALIFA) survey (https://califa.caha.es/) based on observations collected at the Centro Astronómico Hispano Alemán (CAHA) at Calar Alto, operated jointly by the Max-Planck-Institut fűr Astronomie and the Instituto de Astrofísica de Andalucía (CSIC). This work also depended on published data from the NRAO VLA Sky Survey (NVSS). The National Radio Astronomy Observatory and Green Bank Observatory are facilities of the U.S. National Science Foundation operated under cooperative agreement by Associated Universities, Inc. This work depended on published data provided by the PHANGS collaboration and spectrophotometric data published using the Steward Observatory 2.3m (90inch) Bart Bok Telescope at Kitt Peak National Observatory (KPNO), accessed by the NASA Extragalactic Database. This research has made use of the NASA/IPAC Extragalactic Database (NED), which is operated by the Jet Propulsion Laboratory, California Institute of Technology, under contract with the National Aeronautics and Space Administration. This work made use of data from the Galaxy Evolution Explorer, obtained from the MAST data archive at the Space Telescope Science Institute, which is operated by the Association of Universities for Research in Astronomy, Inc., under NASA contract NAS 5–26555. We acknowledge useful discussions with George Helou and Isaac Cheng that helped improve the methodology and discussions contained within this work.
\end{acknowledgments}

%\begin{contribution}
%TDS was responsible for the data collection, analysis, and interpretation. VR was responsible for providing methodological help.
%\end{contribution}

\facilities{CAO:3.5m, Max Planck:2.2m, Du Pont, NRAO:VLA, GALEX, Bok}

\software{\textsc{Astropy}, \citep{2022ApJ...935..167A}, \textsc{emcee} \citep{2013PASP..125..306F}, \textsc{NumPy} \citep{harris2020array}, \textsc{PyPipe3D} \citep{2022NewA...9701895L}, \textsc{scipy} \citep{2020NaMet..17..261V}}


%% For this sample we use BibTeX plus aasjournalv7.bst to generate the
%\clearpage
\appendix

\section{SFG Thermal Fraction Data}
In Table \ref{tab:tab1}, we provide the full list of all galaxies from the CALIFA DR3, PHANGs, and \citetalias{2006ApJS..164...81M} surveys. These are all classified as star forming (or intermediate) based on their nuclear optical spectrum, radio, and IR emission. We show a representative 15 entries across the various surveys in this manuscript. Full table available at \url{https://doi.org/10.5281/zenodo.24541734}.

\startlongtable
\begin{deluxetable*}{lccclccccc}
\tabletypesize{\scriptsize}
\tablewidth{0pt}
\tablecaption{Representative CALIFA+MK06+PHANGS SFG dataset\label{tab:tab1}}
\tablehead{
\colhead{Name} & \colhead{Distance} & \colhead{$cz$} & \colhead{$S_{1.4}$} & \colhead{Survey} & \colhead{$L_{1.4}$} & \colhead{$A(\mathrm{H}\alpha)$} & \colhead{Aperture Corr.} & \colhead{$L_{\mathrm{H}\alpha}$} & \colhead{$\log_{10}(f_{\mathrm{t}})$} \\
\colhead{} & \colhead{(Mpc)} & \colhead{(km\,s$^{-1}$)} & \colhead{(mJy)} & \colhead{} & \colhead{($10^{21}$ W\,Hz$^{-1}$)} & \colhead{(mag)} & \colhead{} & \colhead{($10^{33}$ W)} & \colhead{}
}
\startdata
IC 1528 & 47.84\tablenotemark{a} & 3807 & $12.9 \pm 1.6$ & CT15 & $3.5 \pm 0.4$ & $0.54 \pm 0.03$ & 1.75 & $24.8_{-0.8}^{+0.9}$ & $-1.07_{-0.05}^{+0.06}$ \\
NGC 0023 & 57.02\tablenotemark{a} & 4566 & $74 \pm 3$ & CT15 & $28.5 \pm 1.0$ & $1.14 \pm 0.03$ & 1.25 & $167_{-5}^{+5}$ & $-1.15_{-0.02}^{+0.02}$ \\
UGC 00139 & 47.84\tablenotemark{a} & 3963 & $4.5 \pm 0.6$ & CT15 & $1.23 \pm 0.16$ & $0.46 \pm 0.03$ & 2.11 & $19.5_{-1.1}^{+1.2}$ & $-0.72_{-0.06}^{+0.07}$ \\
IC 0159 & 48.08\tablenotemark{a} & 3915 & $3.1 \pm 0.5$ & This Work & $0.85 \pm 0.13$ & $0.40 \pm 0.06$ & 1.01 & $15.3_{-0.9}^{+0.9}$ & $-0.67_{-0.07}^{+0.07}$ \\
NGC 0693 & 19.43\tablenotemark{a} & 1567 & $52 \pm 2$ & This Work & $2.34 \pm 0.10$ & $0.52 \pm 0.06$ & 1.14 & $6.9_{-0.4}^{+0.4}$ & $-1.45_{-0.03}^{+0.03}$ \\
UGC 10042\tablenotemark{3} & 56.58\tablenotemark{a} & 4224 & $2.2 \pm 0.4$ & This Work & $0.84 \pm 0.15$ & $0.3 \pm 0.2$ & 1.32 & $\gtrsim 12_{-2}^{+3}$ & $\gtrsim -0.76_{-0.11}^{+0.13}$ \\
UGC 00148\tablenotemark{1} & 48.59\tablenotemark{a} & 4213 & $17.8 \pm 1.1$ & This Work & $5.0 \pm 0.3$ & $0.65 \pm 0.07$ & \nodata & $\gtrsim 28.5_{-1.7}^{+1.9}$ & $\gtrsim -1.16_{-0.04}^{+0.04}$ \\
NGC 0681\tablenotemark{1} & 23.37\tablenotemark{a} & 1760 & $30.0 \pm 1.8$ & This Work & $1.96 \pm 0.12$ & $1.01 \pm 0.07$ & \nodata & $\gtrsim 7.2_{-0.5}^{+0.6}$ & $\gtrsim -1.35_{-0.04}^{+0.04}$ \\
UGC 04240\tablenotemark{2} & 118.07\tablenotemark{a*} & 8563 & $5.4 \pm 0.5$ & This Work & $8.9 \pm 0.8$ & \nodata & 1.08 & $\gtrsim 8.4_{-0.9}^{+0.9}$ & $\gtrsim -1.94_{-0.06}^{+0.06}$ \\
IC 1586 & 80.60\tablenotemark{a} & 5821 & $8.0 \pm 0.5$ & MK06 & $6.2 \pm 0.4$ & $0.65 \pm 0.15$ & 1.12 & $49_{-7}^{+9}$ & $-1.02_{-0.07}^{+0.08}$ \\
NGC 1156 & 7.60\tablenotemark{b*} & 375 & $33 \pm 2$ & MK06 & $0.229 \pm 0.017$ & $0.30 \pm 0.15$ & 1.29 & $4.2_{-0.7}^{+0.8}$ & $-0.65_{-0.08}^{+0.08}$ \\
NGC 1036\tablenotemark{1} & 11.00\tablenotemark{a} & 790 & $6.1 \pm 1.1$ & MK06 & $0.088 \pm 0.016$ & $0.37 \pm 0.15$ & \nodata & $\gtrsim 2.0_{-0.3}^{+0.3}$ & $\gtrsim -0.57_{-0.10}^{+0.11}$ \\
NGC 3077\tablenotemark{1} & 3.66\tablenotemark{b} & 14 & $29.0 \pm 1.5$ & MK06 & $0.047 \pm 0.002$ & $0.50 \pm 0.16$ & \nodata & $\gtrsim 1.3_{-0.2}^{+0.2}$ & $\gtrsim -0.47_{-0.08}^{+0.08}$ \\
IC 5273\tablenotemark{2} & 14.18\tablenotemark{c} & 1293 & $40 \pm 3$ & PHANGs & $0.96 \pm 0.07$ & \nodata & \nodata & $\gtrsim 6.79$ & $\gtrsim -1.07_{-0.03}^{+0.04}$ \\
NGC 1300\tablenotemark{2} & 18.99\tablenotemark{c} & 1577 & $33 \pm 3$ & PHANGs & $1.42 \pm 0.13$ & \nodata & \nodata & $\gtrsim 6.67$ & $\gtrsim -1.24_{-0.04}^{+0.04}$ \\
\enddata
\tablecomments{Fifteen representative entries are displayed from the various surveys; all 278 entries are provided in the machine-readable table. The heliocentric velocities $cz$ and 1.4 GHz flux densities $S_{1.4}$ are taken from \cite{2019ApJ...872..148C}. Note that PHANGs H$\alpha$ luminosities are not provided with uncertainties. Radio luminosities are calculated using a $k$-correction with $\alpha=-0.7$ and flux densities provided by \cite{2019ApJ...872..148C}. Flux density uncertainties are derived from the NVSS catalog for single and multi-component sources. Multi-component flux density uncertainties are added in quadrature. Nine sources with \cite{2019ApJ...872..148C} flux densities do not have a matching single or multi-component decompositions in the NVSS catalog, and are approximated with uncertainty $\sigma_f=\sqrt{(0.45^2)+(0.03S_{1.4})^2}\,\mathrm{mJy}$.}
\tablenotetext{a}{These distances are computed using the heliocentric velocities $cz$, converted to the Local Sheet frame using \cite{2008ApJ...676..184T}, and then inputted into the CF4 \citep{Tully_2023} calculator.}
\tablenotetext{*}{These SFGs contain an averaged distance from multiple measurements (either from CF4 or from multiple brightest star measurements).}
\tablenotetext{b}{These SFGs have distances adopted from literature and NED queries. NGC 2337 \citep{2015ApJ...805..144K}, NGC 3077 \citep{McCall2012}, NGC 3274, NGC 3344, NGC 3351, and NGC 4605 \citep{2018ApJS..235...23S}, and NGC 3198 and NGC 4321 \citep{2001ApJ...553...47F} have TRGB or cepheid measurements. NGC 1156 \citep{Kim2012}, NGC 2903 \citep{Drozdovsky2000}, and NGC 6207 \citep{Hendry2006} have brightest-star distances. NGC 3628 uses the cepheid-based group distance in \citetalias{2006ApJS..164...81M}.}
\tablenotetext{c}{These distances are provided by the PHANGs survey \citep{2021MNRAS.501.3621A}.}
\tablenotetext{1}{These sources do not have an aperture correction due to missing GALEX NUV coverage.}
\tablenotetext{2}{These sources do not have an H$\beta$ measurement for internal attenuation correction.}
\tablenotetext{3}{UGC10042 is retained as a lower bound given the inability to interpolate over the bright foreground star, which was instead simply masked. The mask itself includes host galaxy emission.}
\end{deluxetable*}

\section{Thermal Fraction Model}

In this section, we construct a continuous injection cosmic ray loss model that can explain the observed change in thermal fractions with total radio luminosity and hence the non-linearity in $q_{\mathrm{FIR}}$. The thermal radio luminosity $L_{\nu}^{T}\propto C_1Q(H^0)T_e^{0.45}\nu^{-0.1} \propto C_1 \dot{M}_{\star}T_e^{0.45}\nu^{-0.1}$, where $\dot{M}_{\star}$ is the star formation rate, $T_e$ is the electron temperature, and $C_1$ is an overall normalization factor that also accounts for any Lyman continuum losses (e.g., due to escape from the galaxy or absorption by dust) before ionizing neutral hydrogen. We can normalize this relation to the Milky Way (or Milky Way-like galaxies), which has $\dot{M}_{\star}\approx 2\,\mathrm{M_{\odot}\,yr^{-1}}$ \citep[e.g., for a Kroupa initial mass function,][]{Elia_2022} and $L_{1.4}\approx L_{\star}\approx  2\times10^{21}\,\mathrm{W\, Hz^{-1}}$. Based on the observed relation, we find $f_t(L_{\star})\approx 13\%$. Hence, we can normalize the thermal radio emission per unit star formation rate at fixed electron temperature $T_e=10^{4}\,\mathrm{K}$ to be:
\begin{equation}
    \frac{L_{1.4}^{T}}{\dot{M}_{\star}}\approx1.34\times10^{20}\,\mathrm{W\,Hz^{-1}\,M_{\odot}^{-1}\,yr}. 
\end{equation}
This normalization accounts for any potential losses of Lyman continuum photons near $L_{\star}$. We consider the case where there is no significant change in Lyman continuum photon losses relative to $L_{\star}$ across the entire luminosity range and a case where $L_{1.4}^{T}/M_{\star}$ decreases by ${\sim}50\%$ by $\log(L_{1.4}/(\mathrm{W\, Hz^{-1}}))=23$, as suggested by Figure \ref{fig:fig8} comparison B and the \cite{1989ApJ...345..245C} dust law and \cite{10.1093/mnras/stx2613}. This can be represented by adding a total radio-luminosity dependent power law factor to construct a piece-wise function where, for $L_{1.4}\geq 2\times10^{21}\,\mathrm{W\, Hz^{-1}}$,
\begin{equation}
    \frac{L_{1.4}^{T}}{\dot{M}_{\star}}\approx1.34\times10^{20}\left(\frac{L_{1.4}}{L_{\star}}\right)^{-0.16}\,\mathrm{W\,Hz^{-1}\,M_{\odot}^{-1}\,yr};
\end{equation}
the power law index is obtained by taking the difference in the power law index fitted to the observed $q_{\mathrm{FIR}}$ and predicted $q_{\mathrm{FIR}}$ discussed in the previous section.

For the non-thermal emission, under a continuous injection of cosmic rays with spectrum $Q(E)=\kappa E^{-p}$ in the presence of losses, the cosmic ray diffusion equation gives a steady-state cosmic ray electron spectrum of
\begin{equation}
    N(E)=\frac{\kappa E^{-(p-1)}}{(p-1)b(E)},
\end{equation}
where $b(E)=-dE/dt$ is the sum of the various cosmic ray loss powers \cite{2011hea..book.....L}. The non-thermal radio emission is then $L_{\nu}^{\mathrm{NT}}d\nu \propto P_{\mathrm{synch}}N(E)dE$. Assuming that the injection rate of cosmic rays setting the normalization $\kappa$ is directly proportional to the massive star formation rate, we can express the non-thermal radio emission per unit star formation rate to be
\begin{equation}
    \frac{L_{\nu}^{\mathrm{NT}}}{\dot{M_{\star}}}=C_2 \nu^{-p/2}B^{(p-2)/2}\frac{P_{\mathrm{synch}}}{P_{\mathrm{loss}}}.
\end{equation}
Here, $P_{\mathrm{synch}}$ is the synchrotron power per particle, and $P_{\mathrm{loss}}=b(E)$ include ionization, bremsstrahlung, inverse Compton, and synchrotron losses \citep[specified in][]{Murphy_2009}. Following \cite{Murphy_2009}, we include escape into $b(E)$ as an additional effective energy loss mechanism as an approximation. $p\approx 2.6$ is the assumed cosmic ray injection index producing a typical non-thermal spectral index $\alpha_{\mathrm{NT}}=-0.8$ before losses. We have used the fact that a cosmic ray electron producing synchrotron emission at an observed frequency $\nu$ in a magnetic field of strength $B$ will have energy $E\propto \sqrt{\nu/ B}.$ 

To set the overall normalization at 1.4 GHz, we again use the approximate star formation rate and thermal fraction of the Milky Way, following the general procedure of \cite{2026arXiv260602717S}. We additionally assume $B=10\,\mathrm{\mu G}$ for synchrotron losses and $n_{\mathrm{ISM}}=1\,\mathrm{cm^{-3}}$ for ionization/bremsstrahlung losses. For escape losses, we set a fixed $\ell_{\mathrm{esc}}=3\,\mathrm{kpc}$; given the typical $\ell_{\mathrm{synch}} = 1.5\,\mathrm{kpc}$ synchrotron scale height \citep{2018A&A...611A..72K}, and assuming that the cosmic ray energy density is in equipartition with the magnetic field energy density \citep{2016MNRAS.457L..29Y}, $\ell_{\mathrm{esc}}\approx 2\ell_{\mathrm{synch}}$ for a typical non-thermal $\alpha_{\mathrm{NT}}=-0.8$ corresponding to the injection index $p=2.6$ \citep{Murphy_2009}. Finally, we incorporate inverse Compton losses from the CMB approximated to be at $z=0$, and from the interstellar radiation field, setting $U_{\star}=U_B/3$ consistent with observations from \cite{2016MNRAS.457L..29Y}. This produces $P_{\mathrm{synch}}/P_{\mathrm{loss}} \approx 0.30$ and hence a normalization $C_2=2.24\times10^{21}\,\mathrm{W\,Hz^{-1}\,M_{\odot}^{-1}\, yr}.$ Equations B1, B2, and B4 can then be combined to estimate the thermal fraction, normalizing out any star formation rate difference.
\begin{equation}
    f_t = \frac{(L_{1.4}^{T}/\dot{M_{\star}})}{L_{1.4}^{T}/\dot{M}_{\star} + L_{1.4}^{NT}/\dot{M_{\star}}}.
\end{equation}
We approximate thermal and non-thermal emission probing the star formation rate on similar timescales. Estimating thermal fractions for non-$L_{\star}$ galaxies requires a prescription for ISM densities and magnetic field strengths. We consider a scaling $B\propto L_{1.4}^{a}$ with $a$ a free parameter. The ISM density $n_{\mathrm{ISM}}\propto M_{\mathrm{gas}}/R_e^3\propto \Sigma_{\mathrm{gas}}/R_e\propto \Sigma_{\mathrm{SFR}}^{1/1.4}/R_e$ by the Kennicutt-Schmidt relation \citep{1998ApJ...498..541K}. Under the assumption that the magnetic field energy density is related to the supernovae-induced kinetic energy density $B\propto n_{\mathrm{ISM}}^{1/6}\Sigma_{\mathrm{SFR}}^{1/3}$ \citep{2013A&A...556A.142S}. Therefore, under the assumption that the radio continuum sizes of SFGs are weakly dependent on radio luminosity \citep[e.g.,][]{2019A&A...625A.114J}, these relations imply $n_{\mathrm{ISM}}\propto B^{30/19}$, close to the scaling $n\propto B^{2}$ used by \cite{Murphy_2009}. In summary, we specify
\begin{equation}
    B(L_{1.4})=10\left(\frac{L_{1.4}}{L_{\star}}\right)^{a} \,\mathrm{\mu G},
\end{equation}
where $a$ is a free parameter, and
\begin{equation}
    n_{\mathrm{ISM}}=1\,\left(\frac{B(L_{1.4})}{10\,\mathrm{\mu G}}\right)^{30/19}\,\mathrm{cm^{-3}},
\end{equation}
relevant for ionization and bremsstrahlung losses. We retain a fixed cosmic ray escape scale height $\ell=3\,\mathrm{kpc}$ across all SFG radio luminosities given the weak dependence of radio continuum sizes on SFG luminosity/mass \citep{2019A&A...625A.114J}.

\bibliography{sample702}{}
\bibliographystyle{aasjournalv7.1}



\end{document}